\documentclass[twoside]{IEEEtran}
\usepackage{amsmath, amsfonts, amsthm, amssymb}
\usepackage{bbm}
\usepackage{color}

\newcommand{\abs}[1]{{\lvert #1 \rvert}}
\newcommand{\norm}[1]{{\lVert #1 \rVert}}
\newcommand{\set}[1]{{\lbrace{#1}\rbrace}}

\newcommand{\RR}{\mathbb{R}}
\newcommand{\CC}{\mathbb{C}}
\newcommand{\One}{\mathbbm{1}}
\newcommand{\calA}{\mathcal{A}}
\newcommand{\calD}{\mathcal{D}}
\newcommand{\calN}{\mathcal{N}}
\newcommand{\eps}{\varepsilon}
\newcommand{\ct}{\text{ct}}

\newcommand{\column}{\text{column}}

\newcommand{\bA}{\mathbf{A}}
\newcommand{\ba}{\mathbf{a}}
\newcommand{\bD}{\mathbf{D}}

\newcommand{\bF}{\mathbf{F}}
\newcommand{\bvarphi}{\boldsymbol{\varphi}}
\newcommand{\bI}{\mathbf{I}}
\newcommand{\blambda}{\boldsymbol{\lambda}}
\newcommand{\bmu}{\boldsymbol{\mu}}
\newcommand{\bnu}{\boldsymbol{\nu}}
\newcommand{\bPi}{\mathbf{\Pi}}
\newcommand{\bQ}{\mathbf{Q}}
\newcommand{\bs}{\mathbf{s}}
\newcommand{\bt}{\mathbf{t}}
\newcommand{\btheta}{\boldsymbol{\theta}}
\newcommand{\bu}{\mathbf{u}}
\newcommand{\bv}{\mathbf{v}}
\newcommand{\bw}{\mathbf{w}}
\newcommand{\bX}{\mathbf{X}}
\newcommand{\bx}{\mathbf{x}}
\newcommand{\bxi}{\boldsymbol{\xi}}
\newcommand{\bY}{\mathbf{Y}}
\newcommand{\by}{\mathbf{y}}
\newcommand{\bZ}{\mathbf{Z}}
\newcommand{\bzero}{\mathbf{0}}
\newcommand{\bzeta}{\boldsymbol{\zeta}}

\DeclareMathOperator{\EE}{\mathbb{E}}
\newcommand{\E}{{\mathbb E}}

\DeclareMathOperator{\Tr}{tr}
\DeclareMathOperator{\rank}{rank}

\newcommand{\innerprod}[2]{{\left\langle #1, #2 \right\rangle}}

\newtheorem{theorem}{Theorem}[section]
\newtheorem{lemma}[theorem]{Lemma}
\newtheorem{prop}[theorem]{Proposition}

\newtheorem{definition}[theorem]{Definition}
\newtheorem{remark}[theorem]{Remark}

\begin{document}




\title{Phase Retrieval Without\\ Small-Ball Probability Assumptions}


\author{
Felix Krahmer\thanks{F.~Krahmer is with
Research Unit M15,
Department of Mathematics,
Technische Universit\"at M\"unchen
(e-mail: felix.krahmer@tum.de).}
\and \quad
Yi-Kai Liu\thanks{Y.-K.~Liu is with the 
National Institute of Standards and Technology, 
Gaithersburg, MD, USA, 
and also with the
Joint Center for Quantum Information and Computer Science,
University of Maryland
(e-mail: yi-kai.liu@nist.gov).}
\thanks{This paper was presented in part at the 11th International Conference on Sampling Theory and Applications (SampTA 2015), Washington, DC, May 25--29, 2015.}
\thanks{Copyright 2017 IEEE. Personal use of this material is permitted.  However, permission to use this material for any other purposes must be obtained from the IEEE by sending a request to pubs-permissions@ieee.org.}
}

\date{\today}

\maketitle

\begin{abstract}

In the context of the phase retrieval problem, it is known that certain natural classes of measurements, such as Fourier measurements and random Bernoulli measurements, do not lead to the unique reconstruction of all possible signals, even in combination with certain practically feasible random masks.  To avoid this difficulty, the analysis is often restricted to measurement ensembles (or masks) that satisfy a small-ball probability condition, in order to ensure that the reconstruction is unique.  

This paper shows a complementary result:  for random Bernoulli measurements, there is still a large class of signals that \textit{can} be reconstructed uniquely, namely those signals that are \textit{non-peaky}.  In fact, this result is much more general:  it holds for random measurements sampled from \textit{any} subgaussian distribution $\calD$, without any small-ball conditions.  This is demonstrated in two ways:  first, a proof of \emph{stability and uniqueness}, and second, a uniform recovery guarantee for the \emph{PhaseLift algorithm}.  In all of these cases, the number of measurements $m$ approaches the information-theoretic lower bound.  

Finally, for random Bernoulli measurements with erasures, it is shown that PhaseLift achieves uniform recovery of all signals (including peaky ones).  

\end{abstract}

\begin{IEEEkeywords}
Phase retrieval, random measurements, PhaseLift, inverse problems, reconstruction algorithms, sampling methods, convex optimization
\end{IEEEkeywords}


\section{Introduction}

\subsection{Ambiguities in Phase Retrieval}

Phase retrieval is the problem of recovering an unknown vector $\bx \in \CC^n$ from measurements of the form 
\begin{equation}
\label{eqn-one}
y_i = \abs{\ba_i^T\bx}^2 + w_i \quad \text{(for $i=1,\ldots,m$)}, 
\end{equation}
where the vectors $\ba_i \in \CC^n$ are known, and the $w_i \in \RR$ represent additive noise which is unknown.  (The name {\em phase retrieval} refers to the fact that the measurements reveal the magnitudes, but not the phases, of the $\ba_i^T\bx$.)  Phase retrieval has numerous applications including X-ray crystallography \cite{harrison, millane}, astronomy \cite{dainty}, ptychography and coherent diffractive imaging \cite{rodenburg, shechtman}, and quantum state tomography \cite{kueng2014}.  

A typical experiment setup is that one places a detector far from the object being imaged, and the detector records the intensity of the light field, but not its phase. To a first approximation, this situation can be described by phaseless measurements where the $\ba_i$ are Fourier basis vectors. It is well known that such phaseless measurements can give rise to \textit{ambiguities}, in the sense that the solution is not uniquely determined. These ambiguities may include spatial shifts and conjugate inversion. Especially in the one dimensional case, larger classes of ambiguities can arise \cite{BP15}. 

One way to ensure solution uniqueness is to consider a setup with random illuminations, which mathematically corresponds to multiplying each of the image pixels with a randomly chosen factor \cite{fann12}. As argued in \cite{fann12}, a feasible setup is to consider \textit{phase modulations}, where each of these factors lies on the unit circle in the complex plane. Solution uniqueness is then shown in the two-dimensional case, under the assumption that the image has support of rank $2$. These additional assumptions are necessary to exclude simple counterexamples. Namely, any configuration of phase modulations yields measurements 
of the form $\ba_i = (a_{i1}, a_{i2}, \ldots, a_{in})$ where 
\begin{equation}
\label{eqn-same-amplitudes}
\abs{a_{i1}} = \abs{a_{i2}} = \cdots = \abs{a_{in}} = 1. 
\end{equation}
Using such measurements, it is always impossible to distinguish between the vectors $\bx = (1,0,0,\ldots,0)$ and $\mathbf{\tilde{x}} = (0,1,0,\ldots,0)$, since $\abs{\ba_i^T\bx}^2 = 1 = \abs{\ba_i^T\mathbf{\tilde{x}}}^2$.

Motivated by the uniqueness results of \cite{fann12}, a number of follow-up works studied tractable reconstruction algorithms for phase retrieval with random diffraction patterns (cf.~Section~\ref{sec:gauss} below). 
In contrast to \cite{fann12}, however, these works do not restrict the signal class, but rather avoid the ambiguity problems by considering random masks that also vary in \textit{amplitude}, which arguably are more difficult to realize in experiments.


The main goal of this paper is to combine these approaches, studying recoverability under mild assumptions on the class of signals, without requiring measurement vectors of varying amplitude.  
%
 We focus on the simplest class of measurements with property (\ref{eqn-same-amplitudes}), namely random Bernoulli measurements, where the $\ba_i$ are sampled independently and uniformly at random from $\set{1,-1}^n$. We see our results as a proof of concept that measurements of constant amplitude are tractable and expect that this will lay the foundation to study the case of phase modulated Fourier measurement, but we will leave this case for future work.
  
 For random Bernoulli measurements, we show that a surprisingly large class of vectors $\bx$ can be recovered: one can recover all vectors $\bx \in \RR^n$ that are \textit{$\mu$-flat}, in the sense that they satisfy 
\begin{equation}
\label{eqn-lizard}
\norm{\bx}_\infty \leq \mu \norm{\bx}_2, 
\end{equation}
for some constant $\mu \in (0,1)$ that is independent of the dimension $n$. Intuitively, this condition says that $\bx$ is {\em not too peaky}, in the sense that at most a constant fraction of its mass is concentrated on any one coordinate.

Our result for random Bernoulli measurements is a special case of a more general result that applies whenever the measurement vectors $\ba_i$ are sampled independently at random from some subgaussian distribution $\calD$. In the following sections, we will describe this more general setting, and state our results in detail.

\subsection{{Gaussian and Subgaussian Measurements}} \label{sec:gauss}


The scenario in which the measurement vectors $\ba_i$ are chosen at random according to certain distributions has been investigated intensively over the last few years, with the Gaussian distribution being the paradigmatic example. Two main viewpoints have been taken, focusing either on \textit{stable uniqueness}, or on \textit{recovery} via computationally tractable \textit{algorithms}. The first viewpoint asks when $\bx$ is uniquely determined from the measurements \eqref{eqn-one} (up to sign ambiguity and a small reconstruction error resulting from the noise) \cite{EM}. Such stability results are known in rather general settings, where $\bx$ is promised to lie in some known set $T \subset \RR^n$ (for instance, the set of $k$-sparse vectors), and one wants to bound the number of measurements $m$ as a function of some complexity parameter of the set $T$. 

(Note, however, that these stability results \cite{EM} were shown in the real case, where the signal $\bx$ and the measurements $\ba_i$ are in $\RR^n$, rather than $\CC^n$.  For simplicity, in this paper we will likewise focus on the real case.)

The second viewpoint aims at finding tractable algorithms with provable recovery guarantees. A well-known example is PhaseLift, which reduces the problem to one of low-rank matrix recovery, and then solves a convex relaxation \cite{CSV, DH, CL}. In particular, these works show that PhaseLift can recover any vector $\bx \in \CC^n$ using $m = O(n)$ noisy measurements.
  
  The initial works from both of these viewpoints were specific to Gaussian random measurement vectors $\ba_i$. Subsequent work (mainly from the recovery viewpoint) succeeded in partially derandomizing these results, using techniques such as spherical designs \cite{GKK14, kueng2014} and coded diffraction patterns \cite{CLS, GKK}. In addition, stable uniqueness results have been shown for $\ba_i$ chosen from subgaussian distributions, subject to additional assumptions on their small ball probabilities or their fourth moments \cite{EM}. Nonetheless, even these assumptions on the distribution of the $\ba_i$'s are somewhat restrictive. 

It is natural to ask whether these stability analyses and recovery guarantees can be extended to the most straightforward generalization of the Gaussian measurement setup, namely subgaussian measurement vectors $\ba_i$ sampled from a product distribution (i.e., each entry $a_{ij}$ is sampled independently from a subgaussian distribution $\calD$ on $\RR$).  
{However, this opens the door to measurements such as random Bernoulli vectors $\ba_i \in \set{1,-1}^n$, where phase retrieval is not always possible. Previous work on subgaussian phase retrieval has therefore imposed certain restrictions on the distribution of the $\ba_i$, such as small-ball and fourth moment assumptions \cite{EM}, which exclude these pathological cases.}


\subsection{Our Results}

{In this paper we show that a large class of vectors $\bx$ can be recovered uniquely from subgaussian measurements $\ba_i$, \textit{without} imposing any additional conditions on the distribution of the $\ba_i$.  In particular, we show that one can recover all vectors $\bx \in \RR^n$ that are \textit{$\mu$-flat} in the sense of (\ref{eqn-lizard}), where $\mu \in (0,1)$ is a constant that depends on $\calD$, but not on the dimension $n$.}  

In particular, our results apply to Bernoulli measurements $\ba_i$.  Also, note that the $\mu$-flatness requirement does \textit{not} rule out all sparse vectors.  For instance, a vector that has support of size $1/\mu^2$ (i.e., constant size), and that does not have any unusually large entries, will still satisfy equation (\ref{eqn-lizard}), and hence will still be recoverable.  

To some extent, our results are analogous to a recent result on one-bit compressed sensing \cite{ALPY}.  Subgaussian measurements also fail in that context, and this issue can be overcome by restricting to the case of signals which are not too peaky. However, the techniques used there are somewhat different. 

{Our results can also be compared with recent work on phase retrieval using local correlation measurements, which also imposed a ``flatness'' condition on the signal $\bx$ \cite{Iwen15}. However, the flatness condition in that paper is more elaborate than ours, as it depends not only on the magnitudes of the entries in the vector $\bx = (x_1,\ldots,x_n)$, but also on their ordering.\footnote{Essentially, the flatness condition in that paper ensures that the vector $\bx$ does not contain long strings of consecutive 0's. This ensures that, after recovering different pieces of the vector $\bx$ from different sets of local measurements, one can estimate the relative phase-differences among these pieces, in order to ``stitch them together'' and recover $\bx$.}}

We prove three main results.  First, we consider stable uniqueness, as in \cite{EM}.\footnote{A conference version of this first part has appeared in \cite{KL15a}.}  We show that the results of \cite{EM}, on phase retrieval of signals $\bx$ belonging to some set $T \subset \RR^n$, can be generalized to the setting of subgaussian measurements $\ba_i$ (with independent entries $a_{ij}$), provided that all signals in the set $T$ are $\mu$-flat (for some constant $\mu$).  We emphasize that the $\ba_i$ need not satisfy any small-ball or fourth-moment assumptions. 

In particular, we show that the number of measurements $m$ scales with the complexity of the set $T$ in the same way as in \cite{EM}.  For instance, where \cite{EM} showed results on phase retrieval of $k$-sparse vectors in $\RR^n$, we obtain comparable results on phase retrieval of $\mu$-flat $k$-sparse vectors in $\RR^n$.  (Note that a vector can be both $k$-sparse and $\mu$-flat, as long as $k \geq 1/\mu^2$.) 

Second, we prove that the PhaseLift convex program achieves uniform recovery of all $\mu$-flat vectors in $\RR^n$, using $m = O(n)$ subgaussian measurements, in the presence of noise.\footnote{This second part also extends a conference version, which has appeared in \cite{KL15b}.}  This extends the work of Cand\`es and Li, who showed a similar statement for the recovery of all vectors $\bx \in \RR^n$, when the $\ba_i$ are Gaussian distributed \cite{CL}.  

Here, uniform recovery means that, with high probability, a random choice of the $\ba_i$ will allow correct recovery of \textit{all} possible vectors $\bx$; in contrast, a non-uniform guarantee states that for any particular vector $\bx$, with high probability over the choice of the $\ba_i$, $\bx$ will be recovered correctly.  Note that the use of $m = O(n)$ measurements is optimal up to a constant factor.  

Our proof follows a similar approach as Cand\`es and Li \cite{CL}, but we encounter some technical differences, since in our setting the vectors $\ba_i$ do not have the convenient properties of Gaussian random vectors, and at several points we need to exploit the $\mu$-flatness of the vector $\bx$.  

Third, we consider a special class of subgaussian measurements, sometimes called random Bernoulli vectors with erasures.  Roughly speaking,\footnote{In the precise definition, the $a_{ij}$ are multiplied by a normalization factor, so that the distribution has variance 1.} these are random vectors $\ba_i$ whose entries $a_{ij}$ are chosen independently from some symmetric distribution on $\set{1,0,-1}$.  Intuitively, phase retrieval works well with these kinds of measurement vectors, because the presence of zeroes (``erasures'') prevents the pathological behavior that occurs with Bernoulli vectors.  Indeed, this idea was used previously in work on phase retrieval using coded diffraction patterns \cite{CLS,GKK}.

We show that PhaseLift can recover \textit{all} vectors in $\RR^n$ (without any restriction to non-peaky or $\mu$-flat vectors), using $m = O(n)$ random Bernoulli measurements with erasures (using a particular choice $p=2/3$ for the ``erasure probability'').  Our recovery guarantee holds uniformly over all signals in $\RR^n$, with noisy measurements.  

This gives a new example of a class of non-Gaussian measurements where PhaseLift works nearly as well as it does in the case of Gaussian measurements.  We remark that while stability and uniqueness in this setting were previously known (since the $a_{ij}$ satisfy Eldar and Mendelson's fourth-moment conditions) \cite{EM}, there is hardly any previous work involving PhaseLift in this situation.  (The only exception we know is \cite{SKG}, which uses Bernoulli measurements with erasures in an experimental procedure for characterizing linear optical circuits, which have applications in quantum information processing.  Ref.~\cite{SKG} also proves recovery guarantees for PhaseLift with this particular class of measurements, but using a different technique from ours.)

Overall, we see our work as a proof of concept that shows that recovery guarantees are possible for many kinds of random measurements that were not previously considered to be suitable. This is useful in certain situations. For instance, the authors of \cite{SKG} remark that in their optical devices, Bernoulli measurements with erasures can be implemented more easily than Gaussian measurements, because they use fewer levels of quantization. 

An important next step will be to carry over this approach to more realistic and practical scenarios, such as measurements using coded diffraction patterns \cite{CLS, GKK}. Previous work in this area uses measurements that involve random masks that are generated from a very specific distribution. We think it is an interesting question whether one can make a tradeoff, similar to that shown in the present paper, which would allow more flexibility in the choice of masks, at the expense of slightly restricting the class of signals.

\subsection{Outline of the Paper, and Notation}

We introduce some basic definitions in Section \ref{sec:subg}.  We state and prove our results on stable uniqueness in Sections \ref{sec-stab-uniq} and \ref{sec:proofs}.  We then present our results on PhaseLift in Sections \ref{sec-phaselift} through \ref{sec-bernoulli-erasures}.  Finally, we describe some directions for future work in Section \ref{sec-discuss}.

We let $[m]$ denote the set $\set{1,2,\ldots,m}$.  We write vectors in boldface, and matrices in boldface capital letters.  $\norm{\bx}_p$ denotes the $\ell_p$ norm of a vector $\bx$.  $\norm{\bX}_F$ and $\norm{\bX}$ denote the Frobenius  and operator (spectral) norms of a matrix $\bX$, respectively.


\section{Preliminaries}\label{sec:subg}

In this paper, we will consider phaseless measurements of the form (\ref{eqn-one}).  Following \cite{EM}, we will consider the real case, where the signal $\bx$ and the measurement vectors $\ba_i$ are real (rather than complex).  

Also following \cite{EM}, we will suppose that the measurement vectors $\ba_i \in \RR^n$ are sampled independently from some subgaussian distribution.  We recall that a random vector in $\RR^n$ is called \textit{$L$-subgaussian} if all of its  one-dimensional marginals are $L$-subgaussian in the following sense:
\begin{definition} (cf.\ \cite{vershynin})
 A real valued random variable $X$ is {\em subgaussian} with parameter $L$, if for every $u\geq 1$, one has
 \begin{equation*}
  \Pr[|X|\geq Lu] \leq 2 \exp(-u^2/2).
 \end{equation*}
\end{definition}
Here we will consider the (more specific) situation where each $\ba_i$ consists of independent subgaussian entries $a_{ij} \in \RR$, each sampled from some distribution $\calD$.

The main results in \cite{EM} concern measurements $\ba_i$ that are subgaussian and satisfy a \textit{small-ball probability assumption}:  there exists some constant $c > 0$ such that, for all vectors $\bt \in \RR^n$ and for all $\eps > 0$, 
\begin{equation}
\Pr[\abs{\ba_i^T\bt} \leq \eps \norm{\bt}_2] \leq c\eps.  
\end{equation}
In addition, some results are shown in \cite{EM} for measurements that satisfy a fourth-moment condition:  the $a_{ij}$ are symmetric, with variance $\EE(a_{ij}^2) = 1$, and fourth moment $\EE(a_{ij}^4) > 1$.  

In contrast, here we will make no such assumptions on the $\ba_i$.  In particular, our results will hold for Bernoulli measurements $\ba_i$, which are sampled uniformly from the set $\set{1,-1}^n$, and which violate both the small-ball assumption\footnote{This can be seen by setting $\bt = (1,1,0,\ldots,0)$.} and the fourth-moment condition.  

Our results will apply to signals $\bx \in \RR^n$ that are not too peaky, in the following sense.  Let $\mu \in (0,1)$ be a constant that depends on $\calD$, but not on the dimension $n$. 
\begin{definition}
\label{def-mu-flat}
We say that a vector $\bx \in \RR^n$ is $\mu$-flat if it satisfies
\begin{equation}
\norm{\bx}_\infty \leq \mu \norm{\bx}_2, 
\end{equation}
A set $T\subset\RR^n$ is called $\mu$-flat if all its elements are $\mu$-flat.
\end{definition}


\section{Stable Uniqueness}
\label{sec-stab-uniq}

Our first main result concerns {\em stable uniqueness}, and follows the approach taken by Eldar and Mendelson \cite{EM}. That is, the goal will be to find conditions to ensure that if the measurements $\by_1$ and $\by_2$ are close, then the underlying signals $\bx_1$ and $\bx_2$ must also be close (up to sign ambiguity, i.e., either $\bx_1 - \bx_2$ or $\bx_1 + \bx_2$ must be small). 

For conciseness, we write the phaseless measurement operation as 
\begin{equation}
\by=\phi(A\bx)+\bw, 
\end{equation}
where $A \in \RR^{m\times n}$ is the matrix whose $i$'th row is the measurement vector $\ba_i$, 
and $\phi:\: \RR^m \rightarrow \RR^m$ is the function that maps 
\begin{equation}
\phi:\: (s_1, s_2, \dots, s_m) \mapsto (\abs{s_1}^2, \abs{s_2}^2, \ldots, \abs{s_m}^2). 
\end{equation}

\subsection{The Noise-Free Case}

We begin by considering the noise-free case.  Here the notion of stability is formalized in the following definition.  One assumes that the signal $\bx$ lies in some known set $T \subset \RR^n$, in order to address situations where the signal has some known structure, such as sparsity.  
\begin{definition}[Definition 2.3 in \cite{EM}]
 The mapping $\phi(A\bx)$ is \emph{stable} with constant $C>0$ in a set $T \subset \RR^n$ if for every $\bs,\bt\in T$, 
 \begin{equation*}
  \|\phi(A\bs)-\phi(A\bt)\|_1 \geq C \|\bs-\bt\|_2\|\bs+\bt\|_2.
 \end{equation*}
\end{definition}

One then shows that, if the number of measurements $m$ is sufficiently large (with respect to certain parameters that quantify the complexity of the set $T$), then the mapping $\phi(A\bx)$ is stable.  In particular, Eldar and Mendelson \cite{EM} proved results of this type, for several natural choices of the set $T$, where the measurements $\ba_i$ are subgaussian and satisfy small-ball or fourth-moment assumptions.  They showed that stability is achieved with a number of measurements $m$ that is only slightly larger than the information-theoretic lower bound.

We show analogous results, for the same choices of the set $T$ as in \cite{EM}, but restricted to $\mu$-flat vectors with constant $\mu$, and where the measurements $\ba_i$ are subgaussian with independent coordinates, but without any small-ball or fourth-moment assumptions.  In particular, we consider the set $T_\mu\subset \RR^n$ of all $\mu$-flat vectors, and the set $T_{\mu,k}\subset \RR^n$ of all vectors which are both $k$-sparse and $\mu$-flat.  We find that stability holds with a number of measurements $m$ that scales in the same way as in \cite{EM}.
\begin{theorem}\label{thm:example}
 For every $L>0$, there exist constants $c_1,\dots, c_8$ for which the following holds. Let $0 < \mu < \tfrac{1}{2\sqrt{2}}$, and let $T_{\mu,k} \subset T_\mu \subset \RR^n$ be as in the preceding paragraph. 

Consider a random vector $\ba \in \RR^n$ with independent $L$-subgaussian entries $a_j$ with mean zero and unit variance. Let $A\in\RR^{m\times n}$ be a matrix whose rows are independent copies of this vector. Then:
 \begin{itemize}
  \item[(a)] for $u\geq c_1$ and  $m \geq c_2 u^3 \tfrac{ n}{1-8\mu^2}$, one has with probability at least $1-2\exp(- c_3 u^2 n)$ that the mapping $\phi(A\bx)$ is stable with constant $c_4(1 - 8\mu^2)^{1/2}$ in $T_\mu$.
  \item[(b)] for $u\geq c_5$ and  $m \geq c_6 u^3 \tfrac{k \log(en/k)}{1-8\mu^2}$, one has with probability at least $1-2\exp(-c_7 u^2 k \log(en/k))$ that the mapping $\phi(A\bx)$ is stable with constant $c_8(1 - 8\mu^2)^{1/2}$ in $T_{\mu,k}.$
 \end{itemize}
\end{theorem}
To summarize, in these two instances of the phase retrieval problem, most assumptions on the distribution of the measurement vectors $\ba_i$ can be dropped if $\mu$-flatness is introduced as an additional condition on the signal $\bx$, while leaving the other parts of the result unchanged.  As we will see, the proof is quite general, and likely applies to many other instances of the problem.

\subsection{Proof Outline}

We now sketch the proof of Theorem \ref{thm:example}.  This uses the framework introduced in \cite{EM}, with some technical modifications to take advantage of the $\mu$-flatness property of the signals.  First, we define the complexity parameter $\rho_{T,m}$ as follows.  We define $T_+$ and $T_-$ via 
\begin{align*}
 T_-&:=\set{\tfrac{\bs-\bt}{\norm{\bs-\bt}_2}
\::\: \bs,\bt \in T,\,\, \bt \neq -\bs}\\
 T_+&:=\set{\tfrac{\bs+\bt}{\norm{\bs+\bt}_2}
\::\: \bs,\bt \in T,\,\, \bt \neq \bs}
\end{align*}
and then we set $ \rho_{T,m}= \tfrac{E}{\sqrt{m}}+ \tfrac{E^2}{m}$, where 
\begin{equation*}
 E=\max\big(\E \sup_{\bv\in T_-} \sum_{i=1}^n g_i v_i, \E \sup_{\bw\in T_+} \sum_{i=1}^n g_i w_i\big)
\end{equation*}
with $g_i$ independent centered Gaussian random variables of unit variance.

For technical reasons, we will slightly modify the definition of the second complexity parameter $\kappa$ that was used in \cite{EM}. Namely, in our definition of $\kappa$ we restrict to $S^{n-1}$, setting for any $\bv,\bw \in S^{n-1}$ 
\begin{equation}
\label{eqn-kingfisher}
\kappa(\bv,\bw) = \EE| \innerprod{\ba}{\bv} \innerprod{\ba}{\bw} |.
\end{equation}
In this modified notation and restricted to our measurement setup, the main result of \cite{EM} for the noiseless case reads as follows.
\begin{theorem}[Theorem 2.4 in \cite{EM}] \label{thm:EM1}
 For every $L\geq 1$ and $T\subset \RR^n$, there exist constants $c_1, c_2, c_3$  that depend only on $L$ such that the following holds. 

Let $\ba\in\RR^n$ be a random vector with independent, $L$-subgaussian entries with mean zero and unit variance.  Consider a matrix $A\in\RR^{m\times n}$ whose rows are independent copies of this vector. 

Then, for $u\geq c_1$, with probability $\geq 1-2\exp(-c_2 u^2 \min(m, E^2))$, the mapping $\phi(A\bx)$ is stable in $T$ with constant
 \begin{equation}
  C = \inf_{\bs,\bt \in T} \kappa( \tfrac{\bs-\bt}{\norm{\bs-\bt}_2}, \tfrac{\bs+\bt}{\norm{\bs+\bt}_2} ) - c_3 u^3 \rho_{T,m}. 
 \label{eq:stable}
 \end{equation}
 
\end{theorem}

Thus in addition to bounding $\rho_{T,m}$ from above, it suffices to estimate the infimum of $\kappa$ over the set
\begin{equation}
\label{eqn-Tmp}
T_\mp = \set{ ( \tfrac{\bs-\bt}{\norm{\bs-\bt}_2}, \tfrac{\bs+\bt}{\norm{\bs+\bt}_2} ) 
\::\: \bs,\bt \in T,\,\, \bt \neq \bs,\,\, \bt \neq -\bs }.
\end{equation}

As it turns out, this refined infimum allows for sharper bounds when the set under consideration consists of not too peaky vectors.  Our technical contribution consists of a lower bound on $\kappa$, which holds for $\mu$-flat sets:
\begin{prop}\label{prop:main}
For each $L>0$ there exists a constant $c>0$ such that the following holds. Consider a random vector  $\ba$ with independent $L$-subgaussian entries $a_i$ with mean zero and unit variance.  Let $T_\mp$ and $\kappa$ be defined as in equations (\ref{eqn-Tmp}) and (\ref{eqn-kingfisher}). Then if  $T\subset\RR^n$ is $\mu$-flat for some $\mu<\tfrac{1}{2\sqrt{2}}$, one has
\begin{equation}
\label{eqn-water}
\inf_{(\bv,\bw) \in T_{\mp}} \kappa(\bv,\bw) \geq c (1 - 8\mu^2)^{1/2}.
\end{equation}
\end{prop}

\paragraph*{Proof of Theorem~\ref{thm:example}:} We seek to apply Theorem~\ref{thm:EM1}, thus we need to bound the right hand side of \eqref{eq:stable} from below. Applying Proposition~\ref{prop:main} yields a lower bound of $c (1 - 8\mu^2)^{1/2}$ for the first summand. For the second summand, we ignore the $\mu$-flatness, which can only make $\rho_{T,m}$ larger and hence the bound smaller. The resulting setup is exactly the same as in \cite{EM}, so the bounds from Sections 3.3.1 and 3.3.2 in \cite{EM} directly imply 
\[
\rho_{T_\mu, m} \lesssim \sqrt{\tfrac{n}{m}}+\tfrac{n}{m} \quad \text{and} \quad \rho_{T_{\mu,k}, m} \lesssim \sqrt{\tfrac{k \log(en/k)}{m}}+\tfrac{k \log(en/k)}{m}.
\] 
Noting that in both cases, for our choice of $m$, the square root is of leading order, \eqref{eq:stable} yields the result. \qed

\subsection{The Noisy Case}

In \cite{EM}, also an analysis of the case of phase retrieval with noise is presented. The results are technically somewhat more involved. 
It should be noted, however, that again the only place where additional assumptions on the measurement vectors enter is that they ensure a lower bound of $\kappa$. 

A minor difference from the noise-free case is that in this framework, one needs a bound on $\kappa(\tfrac{\bs}{\|\bs\|_2},\tfrac{\bt}{\|\bt\|_2})$, $\bs, \bt\in T$ rather than $\kappa(\tfrac{\bs-\bt}{\|\bs-\bt\|_2},\tfrac{\bs+\bt}{\|\bs+\bt\|_2})$ (both in terms of the definition of $\kappa$ given above, which is slightly different from the one in \cite{EM}). 
We obtain the following bound for the noisy case:
\begin{prop}\label{prop:basic}
 For each $L>0$ there exists a constant $c>0$ such that the following holds. Consider a random vector  $\ba$ with independent $L$-subgaussian entries $a_i$ with mean zero and unit variance.  Let $\kappa$ be defined as in equation (\ref{eqn-kingfisher}). Then if $\bv, \bw\in S^{n-1}$ and at least one of them is $\mu$-flat for some $\mu<\tfrac{1}{\sqrt{2}}$, one has
\begin{equation}
\label{eqn-air}
\kappa(\bv,\bw) \geq c (1 - 2\mu^2)^{1/2}.
\end{equation}
\end{prop}
Taking the infimum over all $\bv=\tfrac{\bs}{\|\bs\|_2}$, $\bw=\tfrac{\bt}{\|\bt\|_2}$ for $\bs,\bt\in T$ yields analogous results to those in \cite{EM} for the noisy case with independent measurement entries, where no small ball probability or moment assumptions are required provided $T$ is $\mu$-flat. Again, the stability constant has an additional factor of $(1-8\mu^2)^{1/2}$. To prove the proposition, the ingredients necessary in addition to the proof in \cite{EM} are exactly the same as in the noise-free case discussed in Theorem~\ref{thm:example} above. We thus refrain from repeating these details.


\section{Lower-bounds on $\kappa$}\label{sec:proofs}

Here we prove the lower-bounds on $\kappa$ that are used to show stable uniqueness.  We will first prove Proposition \ref{prop:basic}, then use it to show Proposition \ref{prop:main}.

\subsection{Proof of Proposition~\ref{prop:basic}}\label{sec:proof1}
By the $\mu$-flatness assumption and as $\bv, \bw\in S^{n-1}$, one has
\begin{equation*}
\norm{\bv}_\infty \leq \mu \text{ or } \norm{\bw}_\infty \leq \mu 
\end{equation*}
and thus 
\begin{equation}\label{eqn-ladybug}
  \sum_{i=1}^n v_i^2 w_i^2 \leq \mu^2 \max(\|\bv\|_2^2, \|\bw\|_2^2)=\mu^2.
  \end{equation}

 Set 
 \begin{equation}
 Z = \innerprod{\ba}{\bv} \innerprod{\ba}{\bw} \label{eqn-zebra}
\end{equation}
 and observe that
\begin{align}
\norm{Z}_{L_2}^2 &= \EE |\innerprod{\ba}{\bv} \innerprod{\ba}{\bw}|^2 \notag \\
& = \EE \sum_{i,j,k,\ell=1}^n a_i a_j a_k a_\ell v_i v_j w_k w_\ell \notag\\
& =  \EE\Big[2\!\!\!\sum_{\substack{i,j=1\\i\neq j}}^n \!\! a_i^2 a_j^2 v_i v_j w_i w_j +\!\!\!\sum_{\substack{i,k=1\\i\neq k}}^n \!\! a_i^2 a_k^2 v_i^2 w_k^2 +\!\! \sum_{i=1}^n a_i^4 v_i^2 w_i^2\Big] \notag\\
&= 1 + 2\innerprod{\bv}{\bw}^2 - 2\sum_{i=1}^n v_i^2 w_i^2 +  \sum_{i=1}^n (\EE a_i^4 - 1)v_i^2 w_i^2 \notag \\
&\geq 1 - 2\mu^2. \label{eqn-platypus}
\end{align}

Here the third equality uses that due to the independence assumption, all summands where an $a_i$ appears in first power have zero mean, so only those terms with two different $a_i$'s appearing as a square or just one $a_i$ appearing in fourth power contribute to the sum. The fourth equality uses that the $a_i$'s are all unit variance, and in the last inequality we use \eqref{eqn-ladybug} as well as the fact that a random variable's fourth moment always dominates its variance.

%
The result now follows tracing exactly the steps of Corollary 3.7 in \cite{EM}. \qed

\subsection{Proof of Proposition~\ref{prop:main}}\label{sec:proof2}

Consider $(\bv,\bw) \in T_{\mp}$. Then by definition, there exist vectors $\bs,\bt \in T$ such that $\bv = \tfrac{\bs-\bt}{\norm{\bs-\bt}_2}$ and $\bw = \tfrac{\bs+\bt}{\norm{\bs+\bt}_2}$.  Using the triangle inequality, and the fact that $\bs,\bt$ are $\mu$-flat, we have that:
\begin{align}
\norm{\bs+\bt}_\infty &\leq \mu (\norm{\bs}_2 + \norm{\bt}_2), \\
\norm{\bs-\bt}_\infty &\leq \mu (\norm{\bs}_2 + \norm{\bt}_2).
\end{align}
Also, using the triangle inequality, 
\begin{align}
\norm{\bs+\bt}_2 &+ \norm{\bs-\bt}_2 \geq \norm{2\bs}_2, \\
\norm{\bs+\bt}_2 &+ \norm{\bs-\bt}_2 \geq \norm{2\bt}_2, 
\end{align}
hence
\begin{equation}
\begin{split}
\norm{\bs}_2 + \norm{\bt}_2 &\leq \norm{\bs+\bt}_2 + \norm{\bs-\bt}_2 \\
&\leq 2\max\set{ \norm{\bs+\bt}_2, \norm{\bs-\bt}_2 }.
\end{split}
\end{equation}
Combining all of the above, we see that at least one of the following inequalities must hold:
\begin{align}
\norm{\bs+\bt}_\infty &\leq 2\mu \norm{\bs+\bt}_2, \\
\norm{\bs-\bt}_\infty &\leq 2\mu \norm{\bs-\bt}_2.
\end{align}
This shows that at least one of $\bs+\bt$ and $\bs-\bt$ is $2\mu$-flat. The result follows by applying Proposition~\ref{prop:basic}. \qed

%
%
%
%


\section{PhaseLift}
\label{sec-phaselift}

Our second main result concerns the PhaseLift algorithm \cite{CSV, DH, CL}.  
PhaseLift (in the noiseless case) is based on a matrix reformulation of the problem  in terms of the rank one matrix $\bX=\bx\bx^T$, namely
\begin{equation}
\begin{split}
\text{find}\quad      &\bX \in \RR^{n\times n} \\
\text{such that}\quad &\bX \succeq 0,\, \rank(\bX) = 1, \\
                      &\Tr(\ba_i\ba_i^T\bX) = y_i \quad (\forall i=1,\ldots,m).
\end{split}
\end{equation}
Then one solves a convex relaxation of this problem:
\begin{equation}\label{eqn-phaselift-noiseless}
\begin{split}
\text{find}\quad      &\bX \in \RR^{n\times n} \text{ that minimizes } \Tr(\bX) \\
\text{such that}\quad &\bX \succeq 0, \\
                      &\Tr(\ba_i\ba_i^T\bX) = y_i \quad (\forall i=1,\ldots,m).
\end{split}
\end{equation}
This yields a solution $\widehat{\bX}$, from which one extracts the leading eigenvector $\hat{\bx}$; one hopes that $\hat{\bx} \approx \bx$.

PhaseLift can be modified to handle noise in different ways; in particular, one can solve the following convex program \cite{CL}:
\begin{equation}\label{eqn-phaselift-noisy}
\begin{split}
\text{find}\quad      &\bX \in \RR^{n\times n} \text{ that minimizes } \sum_{i=1}^m \abs{\Tr(\ba_i\ba_i^T\bX) - y_i} \\
\text{s.t.}\quad &\bX \succeq 0.
\end{split}
\end{equation}

\subsection{Recovery of Non-Peaky Signals from Subgaussian Measurements}

We assume that each measurement vector $\ba_i \in \RR^n$ is chosen by sampling its entries $a_{ij}$ ($j=1,2,\ldots,n$) independently from some subgaussian distribution $\calD$ on $\RR$.  We assume that $\calD$ has mean zero and variance 1.  Also, we let $C_4$ and $C_{\psi_2}$ denote its fourth moment and $\psi_2$-norm, respectively:\footnote{For the definition of the $\psi_2$-norm, see, e.g., \cite{vershynin}.}
\begin{equation}
\label{eqn-D}
\begin{split}
\EE a_{ij} = 0,\qquad
\EE(a_{ij}^2) = 1,\qquad\\
C_4 := \EE(a_{ij}^4) \geq 1,\qquad
C_{\psi_2} := \norm{a_{ij}}_{\psi_2}.
\end{split}
\end{equation}


We are interested in recovering signals that are non-peaky, or $\mu$-flat in the sense of Definition \ref{def-mu-flat}.  We show that, for any choice of $\calD$, there is a constant $\mu \in (0,1)$ such that, for all sufficiently large $n$, PhaseLift achieves uniform recovery of all $\mu$-flat vectors in $\RR^n$, using $m = O(n)$ measurements, in the presence of noise.  

This extends the order-optimal results of Cand\`es and Li \cite{CL} to a larger class of measurement distributions, at the expense of a mild non-peakiness restriction on the class of vectors to be recovered.  We emphasize that there is no dependence of $\mu$ on the dimension $n$. 

%

More precisely, we prove:
\begin{theorem}\label{thm-noisy}
Consider PhaseLift with noisy measurements, as shown in equations (\ref{eqn-one}) and (\ref{eqn-phaselift-noisy}).  Let $\calD$ be any subgaussian distribution on $\RR$, with mean 0, variance 1, and parameters $C_4$ and $C_{\psi_2}$ as shown in equation (\ref{eqn-D}).  Then there exist constants $0 < \mu < 1$ and $\kappa_0 > 1$ such that the following holds.  

For all $n$ sufficiently large, and for all $m \geq \kappa_0 n$, with probability at least $1 - e^{-\Omega(n)}$ (over the choice of the measurement vectors $\ba_i$), PhaseLift achieves uniform recovery of all $\mu$-flat vectors in $\RR^n$:  for all $\mu$-flat vectors $\bx \in \RR^n$, the solution $\widehat{\bX}$ to equation (\ref{eqn-phaselift-noisy}) obeys 
\begin{equation}
\norm{\widehat{\bX} - \bx\bx^T}_F \leq C_0 \frac{\norm{\bw}_1}{m}, 
\end{equation}
where $\norm{\cdot}_F$ is the Frobenius norm, $C_0$ is a universal constant, and $\bw$ is the noise term in equation (\ref{eqn-one}).  If we let $\widehat{\bx}$ be the leading eigenvector of $\widehat{\bX}$, then $\widehat{\bx}$ satisfies the bound 
\begin{equation}
\norm{\widehat{\bx} - e^{i\phi}\bx}_2 \leq C_0 \min\biggl\lbrace \norm{\bx}_2,\, \frac{\norm{\bw}_1}{m\norm{\bx}_2} \biggr\rbrace, 
\end{equation}
for some $\phi \in [0,2\pi]$.
\end{theorem}

Note that here the success probability is $1-e^{-\Omega(n)}$, rather than $1-e^{-\Omega(m)}$ as in \cite{CL}.  This is due to a technical difference in the proof:  in our setting, the sampling operator $\calA$ satisfies the desired injectivity properties with probability $1-e^{-\Omega(n)}$, rather than $1-e^{-\Omega(m)}$.  However, note that the construction of the dual certificate still succeeds with probability $1-e^{\Omega(m)}$, as needed in order to use the union bound over $\RR^n$.  (See the following section for details.)

\subsection{Recovery of Arbitrary Signals from Bernoulli Measurements with Erasures}

Next we consider a special class of subgaussian measurements, namely Bernoulli measurements with erasure probability $p \in [0,1]$.  We let each measurement vector $\ba_i \in \RR^n$ have iid entries chosen from the following distribution:
\begin{equation}\label{eqn-bernoulli-erasures}
a_{ij} = 
\begin{cases}
 1/\sqrt{1-p} &\text{ with probability } (1-p)/2, \\
 0            &\text{ with probability } p, \\
-1/\sqrt{1-p} &\text{ with probability } (1-p)/2.
\end{cases}
\end{equation}
This distribution has the following properties:
\begin{equation}
\begin{split}
\EE a_{ij} = 0, \qquad
&\EE(a_{ij}^2) = 1, \qquad \\
C_4 := \EE(a_{ij}^4) = \tfrac{1}{1-p}, \qquad
&C_{\psi_2} := \norm{a_{ij}}_{\psi_2} \leq \tfrac{1}{\sqrt{1-p}}.
\end{split}
\end{equation}

We consider the phase retrieval problem using these types of measurements, for \textit{arbitrary} signals $\bx \in \RR^n$ (without any $\mu$-flatness assumption).  The basic question of stable uniqueness was already settled by Eldar and Mendelson \cite{EM}:  whenever $p>0$, the distribution of the $a_{ij}$ satisfies their fourth-moment condition, hence stable uniqueness holds.  However, no recovery guarantees using PhaseLift were previously known.  Here we show that, in the particular case where $p = 2/3$, PhaseLift achieves uniform recovery of all vectors in $\RR^n$, using $m = O(n)$ measurements, in the presence of noise.  
\begin{theorem}\label{thm-bernoulli-erasures}
Consider PhaseLift with noisy measurements, as shown in equations (\ref{eqn-one}) and (\ref{eqn-phaselift-noisy}).  Let the $\ba_i \in \RR^n$ be Bernoulli random vectors with erasure probability $p=2/3$, as shown in equation (\ref{eqn-bernoulli-erasures}).  Then there exists a constant $\kappa_0 > 1$ such that the following holds.  

For all $n$ sufficiently large, and for all $m \geq \kappa_0 n$, with probability at least $1 - e^{-\Omega(n)}$ (over the choice of the $\ba_i$), PhaseLift achieves uniform recovery of all vectors in $\RR^n$.  Furthermore, for all $\bx \in \RR^n$, the PhaseLift solution $\widehat{\bX}$ satisfies the same error bounds as in Theorem \ref{thm-noisy}.  
\end{theorem}


\section{Proof Outline}

We begin by describing the proof of Theorem \ref{thm-noisy}.  (We will prove Theorem \ref{thm-bernoulli-erasures} later, in Section \ref{sec-bernoulli-erasures}.)  Our proof uses the same overall strategy as in previous work on PhaseLift \cite{CSV, DH, CL}, though new techniques are needed in several places.

We begin by showing an injectivity property of the sampling operator $\calA:\: \bX \mapsto \bigl( \Tr(\ba_i\ba_i^T\bX) \bigr)_{i\in[m]}$, but while this follows from relatively straightforward arguments in \cite{CL, DH, CSV}, in our setting we need to use more sophisticated arguments due to Eldar and Mendelson \cite{EM} (including a bound on empirical processes from \cite{mendelson}).

To characterize the solution of the PhaseLift convex program, we construct a dual certificate $\bY$, which is similar to that of Cand\`es and Li \cite{CL} (see also similar work by Demanet and Hand \cite{DH}).  Because the $\ba_i$ are no longer Gaussian distributed, our proofs are somewhatmore involved:  we use various fourth-moment estimates, and large-deviation bounds for sums of independent subgaussian and sub-exponential random vectors \cite{vershynin}.


\subsection{Injectivity of the sampling operator}

We define the \textit{sampling operator} $\calA:\: \RR^{n\times n} \rightarrow \RR^m$ as follows:
\begin{equation}
\label{eqn-samplingoperator}
\calA(\bX) = \bigl( \Tr(\ba_i \ba_i^T \bX) \bigr)_{i\in[m]}.
\end{equation}

We will prove that $\calA$ satisfies certain injectivity properties.  First, we need an upper bound on $\norm{\calA(\bX)}_1$, which is a straightforward generalization of the first half of Lemma 2.1 in \cite{CL}, and of Lemma 3.1 in \cite{CSV}:
\begin{lemma}
\label{lem-injectivity1}
Let $\calD$ be as in Theorem~\ref{thm-noisy}.  Then there exist constants $C>0$ and $c>0$ such that the following holds.  

Fix any $0 < \delta < \tfrac{1}{2}$, and assume that $m \geq 20 (C/\delta)^2 n$.  Let $\eps>0$ be the positive root of $\tfrac{1}{4}\delta = \eps^2 + \eps$, that is, $\eps = \tfrac{1}{2} (-1+\sqrt{1+\delta})$.  Then with probability at least $1 - 2e^{-cm\eps^2}$ (over the choice of the $\ba_i$), the sampling operator $\calA$ has the property that:
\begin{equation}
\label{eqn-injectivity1}
\tfrac{1}{m} \norm{\calA(\bX)}_1 \leq (1+\delta) \Tr(\bX), \quad \forall \bX \succeq 0.
\end{equation}
\end{lemma}
The proof is identical to that of Lemma 3.1 in \cite{CSV}, but using bounds on the singular values of random matrices whose rows are independent subgaussian vectors (rather than Gaussian vectors), as in Theorem 5.39 in \cite{vershynin}.

Next we prove a lower bound on $\norm{\calA(\bX)}_1$, which resembles the second half of Lemma 2.1 in \cite{CL}, and Lemma 3.2 in \cite{CSV}.  Our lemma differs from these previous works in that it only applies to matrices $\bX$ that lie in the {\em tangent spaces} $T(\bx_0)$ associated with $\mu$-flat vectors $\bx_0$, rather than all matrices $\bX$ that are symmetric with rank 2.  However, this lemma is sufficient for our purposes.

Formally, fix some $\mu \in (0,1)$, and let $S(\mu)$ be the set of all $\mu$-flat vectors:
\begin{equation}
\label{eqn-setofsignals}
S(\mu) = \set{\bx \in \RR^n \;|\; \norm{\bx}_\infty \leq \mu \norm{\bx}_2}.
\end{equation}
For any vector $\bx_0 \in S$, define $T(\bx_0)$ to be the following subspace of $\RR^{n\times n}$:
\begin{equation}
\label{eqn-tangentspace}
T(\bx_0) = \set{\bX \in \RR^{n\times n} \;|\; \bX = \bx \bx_0^T + \bx_0 \bx^T,\, \bx \in \RR^n}.
\end{equation}
We will prove:
\begin{lemma}
\label{lem-cardinal}
Let $\calD$ be as in Theorem~\ref{thm-noisy}.  Then there exist constants $0 < \mu < 1$, $\kappa_0 > 1$, $c>0$ and $\alpha > 0$ such that the following holds.  

For all $n$ sufficiently large, and for all $m \geq \kappa_0 n$, with probability at least $1 - e^{-cn}$ (over the choice of the measurement vectors $\ba_i$), the sampling operator $\calA$ satisfies:
\begin{equation}
\label{eqn-cardinal}
\tfrac{1}{m} \norm{\calA(\bX)}_1 \geq \alpha \norm{\bX}, \quad 
\forall \bx_0 \in S(\mu) \setminus \set{0},\quad \forall \bX \in T(\bx_0) \setminus \set{0}.
\end{equation}
\end{lemma}

Note that here the constant $\alpha > 0$ may be quite small, in contrast to previous work \cite{CL} where this constant was close to 1.  Because of this difference, we will have to construct a dual certificate $\bY$ that satisfies $\norm{\bY_T}_F \leq \eps_T$ (for a small constant $\eps_T > 0$), rather than $\norm{\bY_T}_F \leq \tfrac{3}{20}$ as in \cite{CL}.  (See Sections \ref{sec-outline-dual-cert} and \ref{sec-outline-error-bound} for details.)

The proof of this lemma is different from \cite{CL, CSV}.  It uses arguments due to Eldar and Mendelson, in particular a Paley-Zygmund argument for lower-bounding an expectation value of the form $\EE \abs{Z}$ (Lemma 3.6 in \cite{EM}), and a bound on the suprema of certain empirical processes (Theorem 2.8 in \cite{EM}, see also \cite{mendelson}).  The proof of this lemma is given in Section \ref{sec-cardinal}.

\begin{remark}
	Note that Lemma~\ref{lem-cardinal} crucially relies on the assumption that $\bx_0$ is $\mu$-flat. For example, let the $\ba_i$ be random Bernoulli measurements in $\set{1,-1}^n$, choose $\bx_0=(1,1, 0, \dots, 0)^T$, and let $\bX= \bx \bx_0^T	+ \bx_0 \bx^T \in T(\bx_0)$ where $\bx = (1,-1, 0,\dots, 0)^T$. This yields $\calA(\bX)=0$ (this follows from~\eqref{penguin} below, as for each of the $\ba_i$, either $\ba_i^T \bx$ or $\ba_i^T \bx_0$ vanishes), so the conclusion of the Lemma does not hold.
\end{remark}	


\subsection{Approximate dual certificates}
\label{sec-outline-dual-cert}

Following \cite{CL}, we will use approximate dual certificates to characterize the solutions to the PhaseLift convex program.  First, we will show that, for any $\mu$-flat vector $\bx_0$, with high probability (over the choice of the measurement vectors $\ba_i$), there exists an approximate dual certificate $\bY$ (for the vector $\bx_0$).  We will then use the union bound over an $\epsilon$-net on the unit sphere $S^{n-1}$ in $\RR^n$, together with a continuity argument (which shows that any vector in $S^{n-1}$ can be approximated by a vector in the $\eps$-net).  This will prove a uniform guarantee:  with high probability (over the choice of the $\ba_i$), for all $\mu$-flat vectors $\bx_0$, there is a dual certificate $\bY$.

For the first claim, we will prove a variant of Lemma 2.3 in \cite{CL}:
\begin{lemma}
\label{lem-dual-cert}
Let $\calD$ be as in Theorem~\ref{thm-noisy}.  Let $0 < \eps_{T^\perp} \leq 1$ and $0 < \eps_T \leq 1$ be any constants.  Then there exist constants $0 < \mu < 1$, $\kappa_0 > 1$, $c>0$ and $B_0 > 0$ such that the following holds.  

For all sufficiently large $n$, and for all $m \geq \kappa_0 n$, let $\bx_0 \in S(\mu)$ be any $\mu$-flat vector, and let $T = T(\bx_0)$ be the tangent space.  Then with probability at least $1 - e^{-cm}$ (over the choice of the measurement vectors $\ba_i$), there exists a {\em dual certificate} $\bY \in \RR^{n\times n}$, which has the form $\bY = \calA^*(\blambda)$, where $\norm{\blambda}_\infty \leq \tfrac{B_0}{m}$, and which satisfies 
\begin{equation}
\norm{\bY_{T^\perp} + 2\bI_{T^\perp}} \leq \eps_{T^\perp}, 
\qquad
\norm{\bY_T}_F \leq \eps_T.  
\end{equation}
\end{lemma}

Here, $\calA^*:\: \RR^m \rightarrow \RR^{n\times n}$ denotes the adjoint of the linear operator $\calA$, $\bY_T$ denotes the projection of $\bY$ onto the subspace $T$, and $\bY_{T^\perp}$ denotes the projection of $\bY$ onto the subspace $T^\perp$, which is the orthogonal complement of $T$ in $\RR^{n\times n}$.

The proof of this lemma is given in Section \ref{sec-dual-cert}.  The dual certificate is constructed in a similar way to \cite{CL}, but the analysis is more involved.  In particular, unlike the Gaussian case studied in \cite{CL}, here the distribution of the vectors $\ba_i$ is not rotationally invariant.  We use vector analogues of Hoeffding's inequality and Bernstein's inequality \cite{vershynin}, combined with fourth-moment estimates which depend on the $\mu$-flatness of the vector $\bx_0$.

From the above lemma, one concludes a uniform guarantee on the existence of dual certificates for all $\mu$-flat vectors $\bx_0$.  To see this, one first constructs a dual certificate for all points $\bx_0$ in an $\epsilon$-net, via a union bound; second, one shows that this dual certificate works well for all points $\bx_0$, using a continuity argument. 
This part of the proof is identical to the proof of Corollary 2.4 in \cite{CL}, so we refrain from repeating the details.

\subsection{Combining the pieces}
\label{sec-outline-error-bound}

We can characterize the solution of the PhaseLift convex program, using the injectivity of the sampling operator $\calA$, and the existence of a dual certificate $\bY$.  We restate the bound proved in Section 2.3 in \cite{CL} (with more general choices for the parameters):
\begin{lemma}
\label{lem-error-bound}
Suppose that the sampling operator $\calA$ satisfies equations (\ref{eqn-injectivity1}) and (\ref{eqn-cardinal}), with parameters $\delta$ and $\alpha$.  Let $\bx_0$ be any $\mu$-flat vector in $\RR^n$, and let $\widehat{\bX}$ be the solution of the PhaseLift convex program with noisy measurements, as in equations (\ref{eqn-one}) and (\ref{eqn-phaselift-noisy}).  

Suppose that there exists a dual certificate $\bY \in \RR^{n\times n}$ that satisfies $\bY_{T^\perp} \preceq -\bI_{T^\perp}$ and $\norm{\bY_T}_F \leq \eps_T$, where $\eps_T < \frac{\alpha}{(1+\delta)\sqrt{2}}$.  Also suppose that $\bY$ has the form $\bY = \calA^*(\blambda)$, where $\norm{\blambda}_\infty \leq \tfrac{B_0}{m}$.  Then $\widehat{\bX}$ must satisfy 
\begin{equation}
\norm{\widehat{\bX} - \bx_0\bx_0^T}_F \leq 2 C_0 \frac{\norm{\bw}_1}{m}, 
\end{equation}
where $C_0 = (1+\eps_T) \bigl( \frac{\alpha}{(1+\delta)\sqrt{2}}-\eps_T \bigr)^{-1} (\frac{1}{1+\delta}+B_0) + B_0$.
\end{lemma}

The proof of this lemma is identical to that in \cite{CL}.  
By combining this with the preceding lemmas, 
we prove Theorem \ref{thm-noisy}.  

\begin{remark}
Theorem \ref{thm-noisy} shows that there exists some constant $\mu > 0$ such that PhaseLift succeeds in recovering all $\mu$-flat vectors.  The specific value of $\mu$ depends on a few factors.  First, in Lemma \ref{lem-cardinal}, in order to show injectivity of the sampling operator, we must have $\mu < 1/\sqrt{2}$ (see equation (\ref{eqn-kappa})).  More importantly, in Lemma \ref{lem-dual-cert}, in order to construct the dual certificate, we need to set $\mu \leq \sqrt{\eps_1/\abs{C_4-3}}$, where $\eps_1 = \tfrac{1}{20} \min\set{\eps_{T^\perp}, \eps_T}$ (see equations (\ref{eqn-momx}) and (\ref{eqn-eps4})).  The specific values of $\eps_{T^\perp}$ and $\eps_T$ are determined by Lemma \ref{lem-error-bound}:  they are $\eps_{T^\perp} = 1$ and $\eps_T < \frac{\alpha}{(1+\delta)\sqrt{2}}$.  The most important factor in determining these values is $\alpha$, which describes the injectivity of the sampling operator in Lemma \ref{lem-cardinal}.  This can be estimated as $\alpha \approx c_1$, where $c_1$ is a numerical constant that comes from the Paley-Zygmund argument from reference \cite{EM} (see equation (\ref{eqn-kappa})).
\end{remark}


\section{Injectivity of the sampling operator}
\label{sec-cardinal}

We now prove Lemma \ref{lem-cardinal}:

For any $\bx_0 \in S(\mu) \setminus \set{0}$, and any $\bX \in T(\bx_0) \setminus \set{0}$, we can write $\bX = \bx \bx_0^T + \bx_0 \bx^T$.  Since all the variables are real, we can write 
\begin{equation} \label{penguin}
\begin{split}
\tfrac{1}{m} \norm{\calA(\bX)}_1
&= \tfrac{1}{m} \sum_{i=1}^m \abs{\Tr(\ba_i \ba_i^T \bX)} 
= \tfrac{2}{m} \sum_{i=1}^m \abs{\ba_i^T \bx} \abs{\ba_i^T \bx_0}.
\end{split}
\end{equation}

Next we reduce to the case where $\norm{\bx}_2 = \norm{\bx_0}_2 = 1$.  Let $S^{n-1}$ denote the unit sphere in $\RR^n$.  For any $\bv \in S^{n-1}$, and any $\bv_0 \in S(\mu) \cap S^{n-1}$, define 
\begin{equation}
\Gamma(\bv,\bv_0) := \tfrac{1}{m} \sum_{i=1}^m \abs{\ba_i^T \bv} \abs{\ba_i^T \bv_0}.
\end{equation}
Then we can write 
\begin{equation}
\begin{split}
\tfrac{1}{m} \norm{\calA(\bX)}_1 
&= 2 \norm{\bx}_2 \norm{\bx_0}_2 \Gamma(\tfrac{\bx}{\norm{\bx}_2}, \tfrac{\bx_0}{\norm{\bx_0}_2}) \\
&\geq \Gamma(\tfrac{\bx}{\norm{\bx}_2}, \tfrac{\bx_0}{\norm{\bx_0}_2}) \norm{\bX},
\end{split}
\end{equation}
where we used the fact that $\norm{\bX} \leq \norm{\bx \bx_0^T} + \norm{\bx_0 \bx^T} \leq 2 \norm{\bx}_2 \norm{\bx_0}_2$.

Furthermore, define 
\begin{equation}
\kappa(\bv,\bv_0) := \EE_{\calA} \Gamma(\bv,\bv_0)
\end{equation}
and 
\begin{equation}
\delta := \sup_{\bv \in S^{n-1},\, \bv_0 \in S(\mu) \cap S^{n-1}} 
\abs{\Gamma(\bv,\bv_0) - \EE_{\calA} \Gamma(\bv,\bv_0)}.
\end{equation}
Then we have 
\begin{equation}
\label{eqn-robin}
\begin{split}
\tfrac{1}{m} \norm{\calA(\bX)}_1 
&\geq \bigl( \kappa(\tfrac{\bx}{\norm{\bx}_2}, \tfrac{\bx_0}{\norm{\bx_0}_2}) - \delta \bigr) \norm{\bX}.
\end{split}
\end{equation}

We now proceed to bound $\kappa(\bv,\bv_0)$, using a version of the Paley-Zygmund argument from Corollary 3.7 in \cite{EM}.  We note that $\kappa(\bv,\bv_0)$ can be written in terms of a single measurement vector $\ba$, as follows:
\begin{equation}
\kappa(\bv,\bv_0) = \EE_\ba \abs{\ba^T \bv} \abs{\ba^T \bv_0} = \EE |Z|,
\label{eqn-petunia}
\end{equation}
where $Z := (\ba^T \bv) (\ba^T \bv_0)$.  
Using \eqref{eqn-platypus}, we have that $\norm{Z}_{L_2} \geq \sqrt{1-2\mu^2}$.  

%

Now fix any $q>2$.  Using the same argument as in Corollary 3.7 in \cite{EM}, we have that $\norm{Z}_{L_q} \leq C_q$, where $C_q < \infty$ depends only on the parameter $q$ and the distribution $\calD$ (and not on the dimension $n$).  Finally, using Lemma 3.6 in \cite{EM}, we get that 
\begin{equation}
\label{eqn-kappa}
\kappa(\bv,\bv_0) = \EE \abs{Z} \geq c_1 \norm{Z}_{L_2} \geq c_1 \sqrt{1-2\mu^2}, 
\end{equation}
where $c_1 > 0$ depends only on the choice of $q$ and 
the distribution $\calD$ (and not on the dimension $n$).  
We assume that $\mu$ is a constant satisfying $\mu < 1/\sqrt{2}$; then $\kappa(\bv,\bv_0)$ is lower-bounded by a constant that is strictly greater than 0.

Next, we will upper-bound $\delta$, by bounding the supremum of an empirical process, using Theorem 2.8 in \cite{EM} (see also \cite{mendelson}).  
Here the empirical process is indexed by two sets $T_1 = S^{n-1}$ and $T_2 = S(\mu) \cap S^{n-1}$.  We define the \textit{diameter} of a set $T \subset \RR^n$ to be 
$d(T) = \sup_{\mathbf{t}\in T} \norm{\mathbf{t}}_2$.
So for the sets $T_1$ and $T_2$, we have $d(T_1) = d(T_2) = 1$.  We define the \textit{Gaussian complexity} of a set $T \subset \RR^n$ to be 
$\ell(T) = \EE_{\mathbf{g}} \sup_{\mathbf{t}\in T} \abs{\mathbf{g}^T\mathbf{t}}$, 
where $\mathbf{g}$ is a Gaussian random vector in $\RR^n$ with mean 0 and covariance matrix $\bI$.  The Gaussian complexities of $T_1$ and $T_2$ can be bounded by $\ell(T_2) \leq \ell(T_1) \leq \sqrt{n}$ (see, e.g., Section 3.3.1 in \cite{EM}).

Now Theorem 2.8 in \cite{EM} implies that 
there exist constants $c_1$, $c_2 > 0$ and $c_3$ (which depend on the distribution $\calD$, but not on the dimension $n$) such that the following holds:  for all $u \geq c_1$, with probability at least $1 - 2\exp(-c_2 u^2 \min\set{m,n})$ (over the choice of the measurement vectors $\ba_i$), we have that 
\begin{equation}
\label{eqn-delta}
\delta \leq c_3 u^3 (\sqrt{\tfrac{n}{m}} + \tfrac{n}{m}).
\end{equation}
Now set $u$ to be constant, and set $m \geq \kappa_0 n$.  Then we have $\delta \leq c_3 u^3 (\tfrac{1}{\sqrt{\kappa_0}} + \tfrac{1}{\kappa_0})$.  By choosing $\kappa_0$ large, we can make $\delta$ arbitrarily small.

Now substitute equations (\ref{eqn-kappa}) and $\ref{eqn-delta}$ into (\ref{eqn-robin}).  This completes the proof of Lemma \ref{lem-cardinal}.  $\square$


\section{Constructing the dual certificate}
\label{sec-dual-cert}

We now prove Lemma \ref{lem-dual-cert}:

\textbf{Part 1:}
We first show some properties of the distribution $\calD$.  We claim that for any $\delta_\ct > 0$, we can define a {\em cutoff radius} $R_\ct > 0$ such that, for any $\bx \in S^{n-1}$, we have 
\begin{equation}
\label{eqn-cutoff}
\Pr[\abs{\ba_i^T\bx} \geq R_\ct] \leq \delta_\ct.
\end{equation}
To see this, use a Hoeffding-type inequality (Prop. 5.10 in \cite{vershynin}):  
\begin{equation}
\Pr[\abs{\ba_i^T\bx} \geq R_\ct] \leq e\cdot \exp(-cR_\ct^2/C_{\psi_2}^2), 
\end{equation}
where $c>0$ is an absolute constant.  Then set $R_\ct$ to be:
\begin{equation}
\label{eqn-Rct}
R_\ct = (\tfrac{1}{c} \ln(\tfrac{e}{\delta_\ct}))^{1/2} C_{\psi_2}.
\end{equation}
We will set $\delta_\ct$ small enough to satisfy various bounds that are needed in the proof, in particular, equations (\ref{eqn-quartz}), (\ref{eqn-mica}) and (\ref{eqn-gravel}).  

We now calculate certain fourth moments of the $\ba_i$, namely $\EE[(\ba_i^T\bx_0)^4]$, $\EE[(\ba_i^T\bx_0)^2 (\ba_i^T\bv)^2]$ and $\EE[(\ba_i^T\bx_0)^3 (\ba_i^T\bv)]$ (where $\bv \in \RR^n$ is a unit vector orthogonal to $\bx_0$).  We show that when the vector $\bx_0$ is $\mu$-flat, these moments have approximately the same values as if $\ba_i$ were a Gaussian random vector.  This is the key fact that allows us to apply the proof techniques from \cite{CL} in this more general setting.

First, let $\bx_0$ be a ``$\mu$-flat'' signal, and assume without loss of generality that $\norm{\bx_0}_2 = 1$.  We consider the 4th moment of $\ba_i^T\bx_0$, and we write 
\begin{equation}
\begin{split}
\EE&[(\ba_i^T\bx_0)^4] = \EE\bigl[ \bigl( \sum_{j=1}^n a_{ij}x_{0j} \bigr)^4 \bigr] \\
&= \sum_j \EE[a_{ij}^4] x_{0j}^4 + 3\sum_{j\neq k} \EE[a_{ij}^2] \EE[a_{ik}^2] x_{0j}^2 x_{0k}^2 \\
&= \sum_j (C_4-3) x_{0j}^4 + 3\sum_{jk} x_{0j}^2 x_{0k}^2 \\
&= \sum_j (C_4-3) x_{0j}^4 + 3.
\end{split}
\end{equation}
In the special case where $\ba_i$ is a Gaussian random vector, we have $C_4 = 3$ and $\EE[(\ba_i^T\bx_0)^4] = 3$.  In the general case, we can use the $\mu$-flatness of $\bx_0$ to show that $\EE[(\ba_i^T\bx_0)^4] \approx 3$, as follows:
\begin{equation}
\label{eqn-momx}
\begin{split}
\abs{\EE[(\ba_i^T\bx_0)^4] - 3} &\leq \abs{C_4-3} \, \norm{\bx_0}_\infty^2 \norm{\bx_0}_2^2 \\
&\leq \abs{C_4-3} \, \mu^2 =: \eps_1.
\end{split}
\end{equation}
Note that we can choose $\mu > 0$ to be small, in order to make $\eps_1$ an arbitrarily small constant.  In particular, let us choose $\mu$ small enough so that:
\begin{equation}
\label{eqn-eps4}
\eps_1 = \tfrac{1}{20} \min\set{\eps_{T^\perp}, \eps_T}.
\end{equation}

Next, let $\bv \in \RR^n$ be a unit vector orthogonal to $\bx_0$, i.e., $\norm{\bv}_2 = 1$ and $\bv^T\bx_0 = 0$.  We calculate the following mixed 4th moment of $\ba_i$:
\begin{equation}
\begin{split}
\EE&[(\ba_i^T\bx_0)^2 (\ba_i^T\bv)^2] \\
&= \sum_j \EE[a_{ij}^4] x_{0j}^2 v_j^2 + \\
&\quad\; \sum_{j\neq k} \EE[a_{ij}^2] \EE[a_{ik}^2] (x_{0j}^2 v_k^2 + x_{0j}x_{0k} v_jv_k + x_{0j}x_{0k} v_kv_j) \\
&= (C_4-3) \sum_j x_{0j}^2 v_j^2 + \sum_{jk} (x_{0j}^2 v_k^2 + 2 x_{0j}x_{0k} v_jv_k) \\
&= (C_4-3) \sum_j x_{0j}^2 v_j^2 + \norm{\bx_0}_2^2 \norm{\bv}_2^2 + 2(\bx_0^T\bv)^2 \\
&= (C_4-3) \sum_j x_{0j}^2 v_j^2 + 1.
\end{split}
\end{equation}
In the special case where $\ba_i$ is a Gaussian random vector, we have $C_4 = 3$ and $\EE[(\ba_i^T\bx_0)^2 (\ba_i^T\bv)^2] = 1$.  In general, we have 
\begin{equation}
\label{eqn-momxv}
\begin{split}
\abs{\EE[(\ba_i^T\bx_0)^2 (\ba_i^T\bv)^2] - 1} &\leq \abs{C_4-3} \, \norm{\bx_0}_\infty^2 \norm{\bv}_2^2 \\
&\leq \abs{C_4-3} \, \mu^2 = \eps_1.
\end{split}
\end{equation}

Finally, we calculate another mixed 4th moment of $\ba_i$:
\begin{equation}
\begin{split}
\EE&[(\ba_i^T\bx_0)^3 (\ba_i^T\bv)] \\
&= \sum_j \EE[a_{ij}^4] x_{0j}^3 v_j + 3 \sum_{j\neq k} \EE[a_{ij}^2] \EE[a_{ik}^2] x_{0j}^2 x_{0k}v_k \\
&= (C_4-3) \sum_j x_{0j}^3 v_j + 3\sum_{jk} x_{0j}^2 x_{0k}v_k \\
&= (C_4-3) \sum_j x_{0j}^3 v_j + 3\norm{\bx_0}_2^2 (\bx_0^T\bv) \\
&= (C_4-3) \sum_j x_{0j}^3 v_j.
\end{split}
\end{equation}
In the special case where $\ba_i$ is a Gaussian random vector, we have $C_4 = 3$ and $\EE[(\ba_i^T\bx_0)^3 (\ba_i^T\bv)] = 0$.  In general, we have 
\begin{equation}
\label{eqn-momx3v}
\begin{split}
\abs{\EE[(\ba_i^T\bx_0)^3 (\ba_i^T\bv)]} 
&\leq \abs{C_4-3} \, \norm{\bx_0}_\infty^2 \norm{\bx_0}_2 \norm{\bv}_2 \\
&\leq \abs{C_4-3} \, \mu^2 = \eps_1.
\end{split}
\end{equation}

\textbf{Part 2:}
We now construct the dual certificate $\bY$, following the same approach as \cite{CL}.  Without loss of generality, we can assume $\norm{\bx_0}_2 = 1$.  We construct $\bY$ as follows:
\begin{equation}
\begin{split}
\bY &:= \sum_{i=1}^m \lambda_i \ba_i \ba_i^T, \qquad \\
\lambda_i &:= \frac{1}{m} \Bigl( (\ba_i^T\bx_0)^2 \, \One[\abs{\ba_i^T\bx_0} \leq R_\ct] - \beta_0 \Bigr),
\end{split}
\end{equation}
where we set 
\begin{equation}
\label{eqn-beta0}
\beta_0 := \EE[(\ba_1^T\bx_0)^4\, \One[\abs{\ba_1^T\bx_0} \leq R_\ct]].  
\end{equation}
This choice of $\beta_0$ ensures that $\EE \bx_0^T\bY\bx_0 = 0$, which will be useful later.  

Note that $\beta_0 \approx \EE[(\ba_1^T\bx_0)^4] \approx 3$, and note that this implies $\abs{\lambda_i} \leq \tfrac{1}{m} (R_\ct^2 + \beta_0) \leq O(\tfrac{1}{m})$.  
In order to make this rigorous, we can bound $\beta_0$ as follows. First define 
\begin{equation}
\label{eqn-beta01}
\begin{split}
\beta_{01} &:= \EE[(\ba_1^T\bx_0)^4] - \beta_0 \\
&=\EE[(\ba_1^T\bx_0)^4\, \One[\abs{\ba_1^T\bx_0} > R_\ct]] \\
&\leq \EE[(\ba_1^T\bx_0)^8]^{1/2} \Pr[\abs{\ba_1^T\bx_0} > R_\ct]^{1/2} \\
&\leq \norm{\ba_1^T\bx_0}_{\psi_2}^4 \sqrt{8}^4 \delta_\ct^{1/2} \\
&\leq C_0^2 C_{\psi_2}^4 \cdot 64 \delta_\ct^{1/2} =: \eps_\ct, 
\end{split}
\end{equation}
where we used the Cauchy-Schwarz inequality and standard properties of subgaussian random variables, and where $C_0$ is some universal constant (see Lemma 5.9 in \cite{vershynin}).  Note that, by choosing $R_\ct$ sufficiently large, we can make $\delta_\ct$, and hence $\eps_\ct$, an arbitrarily small constant.  In particular, let us choose $R_\ct$ large enough so that 
\begin{equation}
\eps_\ct = \tfrac{1}{20} \eps_{T^\perp}.
\end{equation}
Combining with equation (\ref{eqn-momx}), we see that 
\begin{equation}
\abs{\beta_0-3} \leq \eps_1+\eps_\ct \leq \tfrac{1}{10} \eps_{T^\perp}.
\end{equation}

We now show that $\bY$ has the desired properties.  To do this, it is convenient to write $\bY = \bY^{(0)} - \bY^{(1)}$, where 
\begin{equation}
\begin{split}
\bY^{(0)} &:= \frac{1}{m} \sum_{i=1}^m (\ba_i^T\bx_0)^2\, \One[\abs{\ba_i^T\bx_0} \leq R_\ct] \, \ba_i \ba_i^T, \qquad \\
\bY^{(1)} &:= \frac{1}{m} \sum_{i=1}^m \beta_0 \, \ba_i \ba_i^T.
\end{split}
\end{equation}

\textbf{Part 3:}
First, we want to show that $\bY_{T^\perp} + 2\bI_{T^\perp}$ is small.  We can write 
\begin{equation}
\label{eqn-grit}
\norm{\bY_{T^\perp} + 2\bI_{T^\perp}} \leq \norm{\bY_{T^\perp}^{(0)} - \bI_{T^\perp}} + \norm{\bY_{T^\perp}^{(1)} - 3\bI_{T^\perp}}.  
\end{equation}

We begin by showing that $\bY^{(1)}_{T^\perp} - 3\bI_{T^\perp}$ is small --- this is the second term in (\ref{eqn-grit}).  This uses a similar argument to \cite{CL}, with more general bounds for subgaussian random matrices \cite{vershynin}.  We can write 
\begin{equation}
\bY^{(1)} = \tfrac{1}{m} \beta_0 \bA^T\bA, 
\end{equation}
where $\bA$ is a random $m$-by-$n$ matrix whose $i$'th row is the vector $\ba_i^T$.  Note that $\EE \bY^{(1)} = \beta_0 \bI$ and $\EE \bA = \bzero$.  Using standard bounds on the singular values of random matrices with independent subgaussian rows (Lemma 5.39 in \cite{vershynin}), we get that there exist constants $C>0$ and $c>0$ (which depend only on the subgaussian norm of the rows, given by $C_4$), such that for all $t \geq 0$, with probability $\geq 1-2\exp(-ct^2)$, $\bA$ satisfies 
\begin{equation}
\sqrt{m}-C\sqrt{n}-t \leq s_{\min}(\bA) \leq s_{\max}(\bA) \leq \sqrt{m}+C\sqrt{n}+t, 
\end{equation}
where $s_{\max}(\bA)$ and $s_{\min}(\bA)$ denote the largest and smallest singular values of $\bA$.  

We restate this as follows, substituting $t = \eps_0\sqrt{m}$:  for all $\eps_0 \geq 0$, with probability $\geq 1-2\exp(-c\eps_0^2 m)$, $\bA$ satisfies 
\begin{equation}
1-C\tfrac{1}{\sqrt{\kappa_0}}-\eps_0 \leq s_{\min}(\tfrac{\bA}{\sqrt{m}}) \leq s_{\max}(\tfrac{\bA}{\sqrt{m}}) \leq 1+C\tfrac{1}{\sqrt{\kappa_0}}+\eps_0.  
\end{equation}
This implies (using Lemma 5.36 in \cite{vershynin}) that 
\begin{equation}
\norm{\tfrac{1}{m} \bA^T\bA - \bI} \leq 3\max\set{\tfrac{C}{\sqrt{\kappa_0}}+\eps_0, \, (\tfrac{C}{\sqrt{\kappa_0}}+\eps_0)^2}.
\end{equation}
Now set $\eps_0 = \tfrac{1}{90} \eps_{T^\perp}$, and set $\kappa_0$ large enough so that $\tfrac{C}{\sqrt{\kappa_0}} \leq \eps_0$. Then we have 
\begin{equation}
\norm{\tfrac{1}{m} \bA^T\bA - \bI} \leq 3\max\set{2\eps_0, \, (2\eps_0)^2} \leq 6\eps_0 = \tfrac{1}{15} \eps_{T^\perp}.
\end{equation}
This implies 
\begin{equation}
\label{eqn-Y1Tperp}
\begin{split}
\norm{\bY^{(1)}_{T^\perp} - 3\bI_{T^\perp}} 
 &\leq \norm{\bY^{(1)}_{T^\perp} - \beta_0 \bI_{T^\perp}} + \eps_1+\eps_\ct \\
 &\leq \norm{\bY^{(1)} - \beta_0 \bI} + \eps_1 + \eps_\ct \\
 &\leq \tfrac{1}{15} \eps_{T^\perp} \beta_0 + \eps_1 + \eps_\ct \\
 &\leq \tfrac{1}{15} \eps_{T^\perp} (3+\eps_1+\eps_\ct) + \eps_1 + \eps_\ct \\
 &\leq \tfrac{1}{5} \eps_{T^\perp} + (1+\tfrac{1}{15}) (\eps_1+\eps_\ct) \\
 &< (0.31) \, \eps_{T^\perp}.
\end{split}
\end{equation}

Next, we will show that $\bY_{T^\perp}^{(0)} - \bI_{T^\perp}$ is small --- this is the first term in (\ref{eqn-grit}).  This requires a more involved argument.  We can write $\bY_{T^\perp}^{(0)}$ in the form:
\begin{equation}
\bY_{T^\perp}^{(0)} = \frac{1}{m} \sum_{i=1}^m \bxi_i\bxi_i^T, \quad
\bxi_i := (\ba_i^T\bx_0) \One[\abs{\ba_i^T\bx_0} \leq R_\ct] (\bPi_0\ba_i), 
\end{equation}
where $\bPi_0 = \bI - \bx_0\bx_0^T$ is the projector onto the subspace orthogonal to $\bx_0$.  As in \cite{CL}, the random variables $\bxi_i$ are subgaussian, but now there is an added complication, because the $\bxi_i$ may have nonzero mean and may not be isotropic.  We will show that the contributions due to the nonzero means and anisotropy are small, hence the argument from \cite{CL} can be adapted to this situation.


We first shift the $\bxi_i$, in order to get centered random variables $\bzeta_i$ with $\EE \bzeta_i = \bzero$, that is:  
\begin{equation}
\bzeta_i := \bxi_i - \bmu_i, \qquad
\bmu_i := \EE \bxi_i.
\end{equation}
We then define 
\begin{equation}
\bZ := \frac{1}{m} \sum_{i=1}^m \bzeta_i\bzeta_i^T.
\end{equation}
We claim, first, that $\bZ$ is a good approximation to $\bY_{T^\perp}^{(0)}$; second, that the $\bzeta_i$ are approximately isotropic, so $\EE \bZ$ will be close to $\bI_{T^\perp}$; and third, that $\bZ$ is concentrated around its expectation $\EE \bZ$. Finally, we will use these claims to upper-bound $\bY_{T^\perp}^{(0)} - \bI_{T^\perp}$, as follows:
\begin{equation}
\label{eqn-kilo}
\norm{\bY_{T^\perp}^{(0)} - \bI_{T^\perp}}
\leq \norm{\bY_{T^\perp}^{(0)} - \bZ} + \norm{\bZ - \EE \bZ} + \norm{\EE \bZ - \bI_{T^\perp}}
\end{equation}

We begin by upper-bounding $\bY_{T^\perp}^{(0)} - \bZ$ (the first term in (\ref{eqn-kilo})).  The first step is to upper-bound the $\bmu_i$.  We write 
\begin{equation}
\bxi_i = \btheta_i - \bnu_i, 
\end{equation}
where $\btheta_i = (\ba_i^T\bx_0) (\bPi_0\ba_i)$ and $\bnu_i = (\ba_i^T\bx_0) \One[\abs{\ba_i^T\bx_0} > R_\ct] (\bPi_0\ba_i)$.  For any $\bv \in S^{n-1}$ such that $\bv^T\bx_0 = 0$, we have 
\begin{equation}
\EE[\bv^T\btheta_i] = \EE[(\ba_i^T\bx_0) (\bv^T\ba_i)] = \bv^T \EE[\ba_i\ba_i^T] \bx_0 = \bv^T\bx_0 = 0, 
\end{equation}
and using equation (\ref{eqn-momxv}), we have 
\begin{equation}
\begin{split}
\abs{\bv^T \EE \bnu_i} &= \bigl\lvert \EE\bigl[ (\ba_i^T\bx_0) \One[\abs{\ba_i^T\bx_0} > R_\ct] (\bv^T\ba_i) \bigr] \bigr\rvert \\
&\leq \EE[(\ba_i^T\bx_0)^2 (\bv^T\ba_i)^2]^{1/2} \Pr[\abs{\ba_i^T\bx_0} > R_\ct]^{1/2} \\
&\leq \sqrt{1+\eps_1} \sqrt{\delta_\ct}, 
\end{split}
\end{equation}
hence we conclude that $\EE \btheta_i = \bzero$, $\norm{\EE \bnu_i}_2 \leq \sqrt{1+\eps_1} \sqrt{\delta_\ct}$, and 
\begin{equation}
\norm{\bmu_i}_2 \leq \sqrt{1+\eps_1} \sqrt{\delta_\ct}.  
\end{equation}

We can now bound $\bY_{T^\perp}^{(0)} - \bZ$ as follows:
\begin{equation}
\label{eqn-Y0Tperp-Z}
\begin{split}
\norm{\bY_{T^\perp}^{(0)} - \bZ} 
&= \norm{\tfrac{1}{m} \sum_{i=1}^m (\bzeta_i\bmu_i^T + \bmu_i\bzeta_i^T + \bmu_i\bmu_i^T)} \\
&= \norm{(\tfrac{1}{m} \sum_{i=1}^m \bzeta_i) \bmu_1^T + \bmu_1 (\tfrac{1}{m} \sum_{i=1}^m \bzeta_i)^T + \bmu_1\bmu_1^T} \\
&\leq 2 \, \lVert \tfrac{1}{m} \sum_{i=1}^m \bzeta_i \rVert_2 \, \norm{\bmu_1}_2 + \norm{\bmu_1}_2^2 \\
&\leq 2 \, \lVert \tfrac{1}{m} \sum_{i=1}^m \bzeta_i \rVert_2 \, \sqrt{1+\eps_1} \sqrt{\delta_\ct} + (1+\eps_1) \delta_\ct.
\end{split}
\end{equation}

We will bound $\norm{\tfrac{1}{m} \sum_{i=1}^m \bzeta_i}_2$ using a Hoeffding-type inequality for subgaussian random vectors (Lemma \ref{lem-vector-hoeffding}).  Note that the $\bzeta_i$ are independent, centered random variables taking values in $\text{span}(\bx_0)^\perp \subset \RR^n$, and they are subgaussian with norm 
\begin{equation}
\label{eqn-basalt}
\begin{split}
\norm{\bzeta_i}_{\psi_2} &\leq \norm{\bxi_i}_{\psi_2} + \norm{\bmu_i}_{\psi_2} \\
&\leq R_\ct C C_{\psi_2} + \sqrt{1+\eps_1} \sqrt{\delta_\ct} =: K.  
\end{split}
\end{equation}
Using Lemma \ref{lem-vector-hoeffding}, we have that there exists a universal constant $c > 0$ such that, for all $t \geq 0$ and all $0 < \eps < 1$, 
\begin{equation}
\Pr[\lVert \tfrac{1}{m} \sum_{i=1}^m \bzeta_i \rVert_2 \geq \tfrac{1}{1-\eps} t]
\leq e \cdot \exp\bigl( n\ln(\tfrac{3}{\eps}) - \tfrac{ct^2}{K^2} m \bigr).
\end{equation}
Setting $\eps = \tfrac{1}{2}$ and $t = K(1-\eps)$, and recalling that $m \geq \kappa_0 n$, we get that:
\begin{equation}
\Pr[\lVert \tfrac{1}{m} \sum_{i=1}^m \bzeta_i \rVert_2 \geq K]
\leq e \cdot \exp\bigl( \tfrac{\ln(6)}{\kappa_0} m - \tfrac{c}{4} m \bigr).
\end{equation}
By setting $\kappa_0$ is sufficiently large so that $\frac{\ln(6)}{\kappa_0} < \frac{c}{4}$, we get that:
\begin{equation}
\Pr[\lVert \tfrac{1}{m} \sum_{i=1}^m \bzeta_i \rVert_2 \geq K] \leq \exp(-\Omega(m)).
\end{equation}

Thus, with probability $\geq 1 - \exp(-\Omega(m))$, we have that $\lVert \tfrac{1}{m} \sum_{i=1}^m \bzeta_i \rVert_2 \leq K$.  Plugging this into equation (\ref{eqn-Y0Tperp-Z}), using the definition of $K$ in equation (\ref{eqn-basalt}), and setting $R_\ct$ as a function of $\delta_\ct$ as specified in equation (\ref{eqn-Rct}), we get that 
\begin{equation}
\begin{split}
\norm{\bY_{T^\perp}^{(0)} - \bZ} 
&\leq 2 K \sqrt{1+\eps_1} \sqrt{\delta_\ct} + (1+\eps_1) \delta_\ct \\
&\leq 2 R_\ct C C_{\psi_2} \sqrt{1+\eps_1} \sqrt{\delta_\ct} + 3 (1+\eps_1) \delta_\ct \\
&\leq 2 (\tfrac{1}{c} \ln(\tfrac{e}{\delta_\ct}))^{1/2} C_{\psi_2}^2 C \sqrt{1+\eps_1} \sqrt{\delta_\ct} + 3 (1+\eps_1) \delta_\ct.
\end{split}
\end{equation}
This quantity can be made arbitrarily small, by choosing $\delta_\ct$ to be a sufficiently small constant.  In particular, we set $\delta_\ct$ small enough that 
\begin{equation}
\label{eqn-quartz}
\norm{\bY_{T^\perp}^{(0)} - \bZ} \leq \tfrac{1}{10} \eps_{T^\perp}. 
\end{equation}

Next, we show that $\EE \bZ$ is close to $\bI_{T^\perp}$ (corresponding to the third term in (\ref{eqn-kilo})).  First we write 
\begin{equation}
\EE \bZ = \EE \bzeta_i\bzeta_i^T = \EE \bxi_i\bxi_i^T - \bmu_i\bmu_i^T, 
\end{equation}
and note that $\norm{\bmu_i\bmu_i^T} = \norm{\bmu_i}_2^2 \leq (1+\eps_1) \delta_\ct$.  Next we write 
\begin{equation}
\begin{split}
\EE \bxi_i\bxi_i^T 
&= \EE \btheta_i\btheta_i^T - \EE \btheta_i\bnu_i^T - \EE \bnu_i\btheta_i^T + \EE \bnu_i\bnu_i^T \\
&= \EE \btheta_i\btheta_i^T - \EE \bnu_i\bnu_i^T, 
\end{split}
\end{equation}
using the fact that $\btheta_i\bnu_i^T = \bnu_i\btheta_i^T = \bnu_i\bnu_i^T$.  For any $\bv \in S^{n-1}$ such that $\bv^T\bx_0 = 0$, we have 
\begin{equation}
\bv^T \EE[\btheta_i\btheta_i^T] \bv = \EE[(\ba_i^T\bx_0)^2 (\bv^T\ba_i)^2] \in [1-\eps_1, 1+\eps_1], 
\end{equation}
and using the same argument as in equation (\ref{eqn-beta01}), we have 
\begin{equation}
\begin{split}
\abs{\bv^T \EE[\bnu_i\bnu_i^T] \bv} 
&= \bigl\lvert \EE\bigl[ (\ba_i^T\bx_0)^2 \One[\abs{\ba_i^T\bx_0} > R_\ct] (\bv^T\ba_i)^2 \bigr] \bigr\rvert \\
&\leq \EE[(\ba_i^T\bx_0)^4 (\bv^T\ba_i)^4]^{1/2} \Pr[\abs{\ba_i^T\bx_0} > R_\ct]^{1/2} \\
&\leq \EE[(\ba_i^T\bx_0)^8]^{1/4} \EE[(\bv^T\ba_i)^8]^{1/4} \delta_\ct^{1/2} \\
&\leq C_0^2 C_{\psi_2}^4 \cdot 64 \delta_\ct^{1/2} \leq \eps_\ct,
\end{split}
\end{equation}
hence we conclude that $\norm{\EE \btheta_i\btheta_i^T - \bI_{T^\perp}} \leq \eps_1$ and $\norm{\EE \bnu_i\bnu_i^T} \leq \eps_\ct$.  Combining these pieces, and setting $\delta_\ct$ sufficiently small, we get that 
\begin{equation}
\label{eqn-mica}
\begin{split}
\norm{\EE \bZ - \bI_{T^\perp}} &\leq \eps_1 + \eps_\ct + (1+\eps_1)\delta_\ct \\
 &\leq \tfrac{1}{10} \eps_{T^\perp} + (1+\eps_1)\delta_\ct \\
 &\leq \tfrac{1}{5} \eps_{T^\perp}.
\end{split}
\end{equation}

Finally, we will upper-bound $\bZ - \EE \bZ$ (the second term in (\ref{eqn-kilo})).  To accomplish this, we will whiten the random variables $\bzeta_i$, to make them isotropic.  We consider the covariance matrix of the $\bzeta_i$, and we compute its spectral decomposition, 
\begin{equation}
\EE \bZ = \EE \bzeta_i\bzeta_i^T =: \bQ\bD\bQ^T. 
\end{equation}
Then we transform each $\bzeta_i$ to get an isotropic random variable $\bvarphi_i$:
\begin{equation}
\bvarphi_i := \bD^{-1/2} \bQ^T \bzeta_i, 
\end{equation}
where $\bQ$, $\bD$ and $\bD^{-1/2}$ are understood to act on the subspace orthogonal to $\bx_0$.  We also define 
\begin{equation}
\bF := \frac{1}{m} \sum_{i=1}^m \bvarphi_i\bvarphi_i^T,
\end{equation}
so we have $\EE \bF = \bI_{T^\perp}$.  We can then upper-bound $\bZ - \EE \bZ$ as follows:
\begin{equation}
\begin{split}
\norm{\bZ - \EE \bZ} &= \norm{\bQ\bD^{1/2} (\bF-\bI_{T^\perp}) \bD^{1/2}\bQ^T} \\
&\leq \norm{\bD} \, \norm{\bF-\bI_{T^\perp}} \\
&\leq (1 + \tfrac{1}{5} \eps_{T^\perp}) \, \norm{\bF-\bI_{T^\perp}}, 
\end{split}
\end{equation}
where we used equation (\ref{eqn-mica}).  

We can now bound $\bF = \frac{1}{m} \sum_{i=1}^m \bvarphi_i\bvarphi_i^T$ as follows.  Note that the $\bvarphi_i$ are subgaussian, with norm 
\begin{equation}
\begin{split}
\norm{\bvarphi_i}_{\psi_2}
&\leq \norm{\bD^{-1/2}} \norm{\bzeta_i}_{\psi_2} \\
&\leq \frac{\norm{\bxi_i}_{\psi_2} + \norm{\bmu_i}_{\psi_2}}{\sqrt{1 - \tfrac{1}{5} \eps_{T^\perp}}} \\
&\leq \frac{R_\ct C C_{\psi_2} + \sqrt{1+\eps_1} \sqrt{\delta_\ct}}{\sqrt{\tfrac{4}{5}}}, 
\end{split}
\end{equation}
which is constant with respect to $n$.  We bound $\bF$ using the same argument that led to equation (\ref{eqn-Y1Tperp}).  We get that there exist constants $C>0$ and $c>0$ (which depend only on $\norm{\bvarphi_i}_{\psi_2}$), such that for all $\eps_0 \geq 0$, with probability $\geq 1-2\exp(-c\eps_0^2 m)$, $\bF$ satisfies 
\begin{equation}
\label{eqn-F}
\norm{\bF - \bI_{T^\perp}} \leq 6\eps_0.
\end{equation}
In particular, we can set $\eps_0 = \tfrac{1}{30} \eps_{T^\perp}$.  This implies the following bound on $\bZ - \EE \bZ$:
\begin{equation}
\label{eqn-silica}
\begin{split}
\norm{\bZ - \EE \bZ} &\leq \tfrac{1}{5} \eps_{T^\perp} (1 + \tfrac{1}{5} \eps_{T^\perp}) \\
&\leq (0.24) \, \eps_{T^\perp}.
\end{split}
\end{equation}

Combining equations (\ref{eqn-kilo}), (\ref{eqn-quartz}), (\ref{eqn-mica}) and (\ref{eqn-silica}), we see that with probability $\geq 1-e^{-\Omega(m)}$, we have 
\begin{equation}
\label{eqn-grate}
\begin{split}
\norm{\bY_{T^\perp}^{(0)} - \bI_{T^\perp}} 
&\leq \tfrac{1}{10} \eps_{T^\perp} + \tfrac{1}{5} \eps_{T^\perp} + (0.24) \, \eps_{T^\perp} \\
&= (0.54) \, \eps_{T^\perp}.
\end{split}
\end{equation}

Finally, combining equations (\ref{eqn-grit}), (\ref{eqn-Y1Tperp}) and (\ref{eqn-grate}), we see that with probability $\geq 1-e^{-\Omega(m)}$, 
\begin{equation}
\norm{\bY_{T^\perp} + 2\bI_{T^\perp}} \leq (0.31) \, \eps_{T^\perp} + (0.54) \, \eps_{T^\perp}
= (0.85) \, \eps_{T^\perp},
\end{equation}
as desired.

\textbf{Part 4:}
We now want to bound $\bY_T$.  Following \cite{CL}, we can write 
\begin{equation}
\label{eqn-YTF2}
\norm{\bY_T}_F^2 = \abs{\bx_0^T\bY\bx_0}^2 + 2\norm{\bPi_0\bY\bx_0}_2^2, 
\end{equation}
where $\bPi_0 = \bI - \bx_0\bx_0^T$ is the projector onto $\text{span}(\bx_0)^\perp$, and we assume (without loss of generality) that $\norm{\bx_0}_2 = 1$.  We will bound each term in equation (\ref{eqn-YTF2}) separately.


For the first term in (\ref{eqn-YTF2}), we use same argument as in \cite{CL}.
We can write 
\begin{equation}
\bx_0^T\bY\bx_0 = \frac{1}{m} \sum_{i=1}^m \xi_i, 
\end{equation}
where $\xi_i := (\ba_i^T\bx_0)^4 \One[\abs{\ba_i^T\bx_0} \leq R_\ct] - \beta_0 (\ba_i^T\bx_0)^2$.  Note that $\EE \xi_i = 0$, by our choice of $\beta_0$ in equation (\ref{eqn-beta0}).  Furthermore, $\xi_i$ is a sub-exponential random variable, since the first term is bounded by $R_\ct^4$, and the second term is the square of a subgaussian random variable.  In particular, we can write 
\begin{equation}
\begin{split}
\norm{\xi_i}_{\psi_1} &\leq R_\ct^4 + \beta_0 \norm{(\ba_i^T\bx_0)^2}_{\psi_1} \\
&\leq R_\ct^4 + 2\beta_0 \norm{\ba_i^T\bx_0}_{\psi_2}^2 \\
&\leq R_\ct^4 + 2\beta_0 C^2 C_{\psi_2}^2 \\
&=: K, 
\end{split}
\end{equation}
where we used Lemmas 5.14 and 5.24 in \cite{vershynin}, and $C$ is some universal constant.  Thus we can use a Bernstein-type inequality (Cor. 5.17 in \cite{vershynin}) to bound $\bx_0^T\bY\bx_0$.  We get that there exists a universal constant $c>0$ such that, for all $\eps \geq 0$, 
\begin{equation}
\Pr[\abs{\bx_0^T\bY\bx_0} \geq \eps] \leq 2\exp(-c\min\set{ \tfrac{\eps^2}{K^2}, \, \tfrac{\eps}{K} }m).
\end{equation}
Now set $\eps = \tfrac{1}{2} \eps_T$.  Then, with probability $\geq 1 - \exp(-\Omega(m))$, we have 
\begin{equation}
\label{eqn-Ycorner}
\abs{\bx_0^T\bY\bx_0} < \tfrac{1}{2} \eps_T.
\end{equation}

For the second term in (\ref{eqn-YTF2}), we need to use a different argument.  The proof in \cite{CL} uses the fact that $\ba_i^T\bx_0$ and $\bPi_0\ba_i$ are \textit{independent} random variables, when $\ba_i$ is sampled from a spherical Gaussian distribution; but this no longer holds in our setting.  Instead, we give a more general argument.  We write 
\begin{equation}
\bPi_0\bY\bx_0 = \frac{1}{m} \sum_{i=1}^m \bs_i - \bt_i, 
\end{equation}
where $\bs_i := (\ba_i^T\bx_0)^3 \One[\abs{\ba_i^T\bx_0} \leq R_\ct] (\bPi_0\ba_i)$, and $\bt_i := \beta_0 (\ba_i^T\bx_0) (\bPi_0\ba_i)$.  We note that $\bs_i$ is a subgaussian random vector and $\bt_i$ is a sub-exponential random vector.  We then use a Bernstein-type inequality (for vectors rather than scalars) to bound $\bPi_0\bY\bx_0$.

To make this precise, we define centered random variables $\bu_i = \bs_i - \bt_i - \EE\bs_i + \EE\bt_i$, and we write 
\begin{equation}
\label{eqn-copper}
\bPi_0\bY\bx_0 = \frac{1}{m} \sum_{i=1}^m \bu_i + \frac{1}{m} \sum_{i=1}^m (\EE\bs_i - \EE\bt_i).
\end{equation}

We begin by calculating $\EE \bs_i$ and $\EE \bt_i$.  Let us write $\bs_i$ as a difference of two terms, 
\begin{equation}
\bs_i = \bs_{i0} - \bs_{i1}, 
\end{equation}
where we define $\bs_{i0} := (\ba_i^T\bx_0)^3 (\bPi_0\ba_i)$ and $\bs_{i1} := (\ba_i^T\bx_0)^3 \One[\abs{\ba_i^T\bx_0} > R_\ct] (\bPi_0\ba_i)$.  Now consider any vector $\bv \in S^{n-1}$ such that $\bv^T\bx_0 = 0$.  We have the following bounds:
\begin{equation}
\abs{\bv^T (\EE\bs_{i0})} = \abs{\EE[(\ba_i^T\bx_0)^3 (\bv^T\ba_i)]} \leq \eps_1, 
\end{equation}
where we used equation (\ref{eqn-momx3v}); 
\begin{equation}
\begin{split}
\abs{\bv^T (\EE\bs_{i1})} &= \abs{\EE[(\ba_i^T\bx_0)^3 (\bv^T\ba_i) \One[\abs{\ba_i^T\bx_0} > R_\ct]]} \\
&\leq \EE[(\ba_i^T\bx_0)^6 (\bv^T\ba_i)^2]^{1/2} \, \Pr[\abs{\ba_i^T\bx_0} > R_\ct]^{1/2} \\
&\leq \EE[(\ba_i^T\bx_0)^{12}]^{1/4} \, \EE[(\bv^T\ba_i)^4]^{1/4} \, \delta_\ct^{1/2} \\
&\leq (\norm{\ba_i^T\bx_0}_{\psi_2} \sqrt{12})^3 \, (\norm{\bv^T\ba_i}_{\psi_2} \sqrt{4}) \, \delta_\ct^{1/2} \\
&\leq (CC_{\psi_2})^4 \, (48\sqrt{3}) \, \delta_\ct^{1/2},
\end{split}
\end{equation}
where we used standard properties of subgaussian random variables, with $C$ a universal constant; and 
\begin{equation}
\begin{split}
\bv^T (\EE\bt_i) &= \EE[\beta_0 (\bv^T\ba_i) (\ba_i^T\bx_0)] \\
&= \beta_0 \bv^T \EE[\ba_i\ba_i^T] \bx_0 = \beta_0 \bv^T\bx_0 = 0.
\end{split}
\end{equation} 
Thus we conclude that $\EE\bs_i$ is bounded as follows:  
\begin{equation}
\label{eqn-gravel}
\begin{split}
\norm{\EE\bs_i}_2 &\leq \norm{\EE\bs_{i0}}_2 + \norm{\EE\bs_{i1}}_2 \\
&\leq \eps_1 + (CC_{\psi_2})^4 \, (48\sqrt{3}) \, \delta_\ct^{1/2} \\
&\leq \tfrac{1}{20} \eps_T + \tfrac{1}{5} \eps_T \\
&= \tfrac{1}{4} \eps_T,
\end{split}
\end{equation}
where we chose $\delta_\ct$ sufficiently small; and we conclude that 
\begin{equation}
\label{eqn-gravel2}
\EE\bt_i = \bzero.
\end{equation}

Next, we note that $\bs_i$ and $\bt_i$ are sub-exponential random vectors, whose norms are bounded by:
\begin{equation}
\begin{split}
\norm{\bs_i}_{\psi_1} &= \sup_{\bv \in S^{n-1}} \norm{\bv^T\bs_i}_{\psi_1} 
\leq \sup_{\bv \in S^{n-1}} \norm{\bv^T\bs_i}_{\psi_2} \\
&\leq \sup_{\bv \in S^{n-1}} R_\ct^3 \norm{\bv^T\bPi_0\ba_i} 
\leq R_\ct^3 C C_{\psi_2},
\end{split}
\end{equation}
and 
\begin{equation}
\begin{split}
\norm{\bt_i}_{\psi_1} &= \sup_{\bv \in S^{n-1}} \norm{\bv^T\bt_i}_{\psi_1} \\
&\leq \sup_{\bv \in S^{n-1}} 2\beta_0 \norm{\ba_i^T\bx_0}_{\psi_2} \norm{\bv^T\bPi_0\ba_i}_{\psi_2} \\
&\leq 2 (3+\eps_1+\eps_\ct) (C C_{\psi_2})^2, 
\end{split}
\end{equation}
where $C$ is a universal constant, and we used standard properties of sub-exponential random variables. 
\footnote{In particular, note that for any two scalar-valued random variables $X$ and $Y$, 
\begin{equation}
\begin{split}
\norm{XY}_{\psi_1} &= \sup_{p\geq 1} \tfrac{1}{p} \EE[\abs{XY}^p]^{1/p} \\
&\leq \sup_{p\geq 1} \tfrac{1}{p} \EE[\abs{X}^{2p}]^{1/2p} \EE[\abs{Y}^{2p}]^{1/2p} \\
&\leq 2 \sup_{p\geq 1} \tfrac{1}{\sqrt{2p}} \EE[\abs{X}^{2p}]^{1/2p} \sup_{q\geq 1} \tfrac{1}{\sqrt{2q}} \EE[\abs{Y}^{2q}]^{1/2q} \\
&\leq 2 \norm{X}_{\psi_2} \norm{Y}_{\psi_2}.
\end{split}
\end{equation}
}

This implies that the $\bu_i$ are sub-exponential, with norm 
\begin{equation}
\begin{split}
\norm{\bu_i}_{\psi_1} &\leq \norm{\bs_i}_{\psi_1} + \norm{\bt_i}_{\psi_1} + \norm{\EE\bs_i}_2 \\
&\leq R_\ct^3 C C_{\psi_2} + 2 (3+\eps_1+\eps_\ct) (CC_{\psi_2})^2 + \eps_{\column1} \\
&=: K.
\end{split}
\end{equation}
We will now bound $\tfrac{1}{m} \sum_{i=1}^m \bu_i$, using a Bernstein-type inequality (Lemma \ref{lem-vector-bernstein}).  We get that, for any $t \geq 0$ and any $0 < \eps < 1$, 
\begin{equation}
\Pr\bigl[ \lVert \tfrac{1}{m} \sum_{i=1}^m \bu_i \rVert_2 \geq \tfrac{1}{1-\eps} t \bigr] 
\leq 2\exp\bigl( n\ln(\tfrac{3}{\eps}) - c\min\set{ \tfrac{t^2}{K^2}, \, \tfrac{t}{K} }m \bigr), 
\end{equation}
where $c>0$ is a universal constant.  We set $\eps = \tfrac{1}{2}$, substitute $t \mapsto t(1-\eps)$, and recall that $m \geq \kappa_0 n$; this gives 
\begin{equation} 
\Pr\bigl[ \lVert \tfrac{1}{m} \sum_{i=1}^m \bu_i \rVert_2 \geq t \bigr] 
\leq 2\exp\bigl( \tfrac{1}{\kappa_0} \ln(6) m - c\min\set{ \tfrac{t^2}{4K^2}, \, \tfrac{t}{2K} }m \bigr).
\end{equation} 
Now set $t = \tfrac{1}{4} \eps_T$.  Then there exists some (sufficiently large) constant $\kappa_0 > 0$ such that 
\begin{equation}
\label{eqn-pebble}
\Pr\bigl[ \lVert \tfrac{1}{m} \sum_{i=1}^m \bu_i \rVert_2 \geq \tfrac{1}{4} \eps_T \bigr] 
\leq \exp(-\Omega(m)).
\end{equation}

Finally we combine equations (\ref{eqn-copper}), (\ref{eqn-gravel}), (\ref{eqn-gravel2}) and (\ref{eqn-pebble}) to get a bound on $\bPi_0\bY\bx_0$:  with probability $\geq 1 - \exp(-\Omega(m))$, we have 
\begin{equation}
\label{eqn-Ycolumn}
\norm{\bPi_0\bY\bx_0}_2 \leq \tfrac{1}{4} \eps_T + \tfrac{1}{4} \eps_T = \tfrac{1}{2} \eps_T.
\end{equation}

Combining equations (\ref{eqn-YTF2}), (\ref{eqn-Ycorner}) and (\ref{eqn-Ycolumn}), we get the following bound on $\bY_T$:  with probability $\geq 1-e^{-\Omega(m)}$, 
\begin{equation}
\norm{\bY_T}_F^2 \leq \tfrac{1}{4} \eps_T^2 + 2 \cdot \tfrac{1}{4} \eps_T^2 = \tfrac{3}{4} \eps_T^2.
\end{equation}
This completes the proof.  $\square$


\section{Proof of Theorem \ref{thm-bernoulli-erasures}}
\label{sec-bernoulli-erasures}

To prove Theorem \ref{thm-bernoulli-erasures}, we follow the same strategy used to prove Theorem \ref{thm-noisy}, as described in the preceding sections.  Now the measurement vectors $\ba_i \in \RR^n$ are Bernoulli random vectors with erasures, which are a special case of the subgaussian random vectors considered previously; hence most of the proof goes through in the same way as before.  

The present situation is different, however, in that the signal $\bx \in \RR^n$ is no longer assumed to be $\mu$-flat.  This affects Lemma \ref{lem-cardinal} (injectivity of the sampling operator) and Lemma \ref{lem-dual-cert} (construction of the dual certificate).  We claim that, if the $\ba_i$ are Bernoulli random vectors with erasure probability $p=2/3$, then we can set $\mu=1$ in both of these lemmas, so that both lemmas now apply to all $\bx$ (as the $\mu$-flatness condition is trivially satisfied).  

In the original proofs of these two lemmas, the $\mu$-flatness conditions are used via equations (\ref{eqn-platypus}), (\ref{eqn-momx}), (\ref{eqn-momxv}) and (\ref{eqn-momx3v}), which bound certain 4th moments of the $\ba_i$, involving projections onto vectors $\bv$ that are $\mu$-flat.  We claim that similar bounds hold in our new situation (where the $\ba_i$ are Bernoulli random vectors with erasure probability $p=2/3$, and we allow the vector $\bv$ to be arbitrary).  

To see this, note that when $p=2/3$, the $a_{ij}$ have mean $\EE a_{ij} = 0$, variance $\EE(a_{ij}^2) = 1$ and fourth moment $C_4 := \EE(a_{ij}^4) = \tfrac{1}{1-p} = 3$, which are the same as the moments of the Gaussian distribution.  Intuitively, this means that the $\ba_i$ behave like Gaussian random vectors, which are rotationally symmetric in $\RR^n$.  Thus, there is no pathological behavior in the 4th moments of the $\ba_i$, when one projects onto a direction $\bv$ that is not $\mu$-flat.  

To make this rigorous, we can redo the calculation of equation (\ref{eqn-platypus}) as follows:  for any $\bv,\bw \in S^{n-1}$, 
\begin{equation}
\begin{split}
\EE[ &(\ba_i^T \bv)^2 (\ba_i^T \bw)^2 ] \\
&= C_4 \sum_{i=1}^n v_i^2 w_i^2 + \sum_{i\neq j} v_i^2 w_j^2 + 2 \sum_{i\neq j} (v_i w_i) (v_j w_j) \\
&= (C_4-3) \sum_{i=1}^n v_i^2 w_i^2 + \norm{\bv}_2^2 \norm{\bw}_2^2 + 2 (\bv^T\bw)^2 \\
&= 1 + 2 (\bv^T\bw)^2.
\end{split}
\end{equation}
Note that this bound holds for \textit{arbitrary} unit vectors $\bv$ and $\bw$, which need not be $\mu$-flat.  This bound serves the same purpose as equation (\ref{eqn-platypus}).  When we set $\bv = \bw$, this bound serves the same purpose as equation (\ref{eqn-momx}).  When we choose $\bv$ and $\bw$ that are orthogonal, this bound serves the same purpose as equation (\ref{eqn-momxv}).  

Finally, we can redo the calculation of equation (\ref{eqn-momx3v}) as follows:  for any $\bv,\bw \in S^{n-1}$, 
\begin{equation}
\begin{split}
\EE&[(\ba_i^T\bv)^3 (\ba_i^T\bw)] \\
&= C_4 \sum_j v_j^3 w_j + 3 \sum_{j\neq k} v_j^2 v_k w_k \\
&= (C_4-3) \sum_j v_j^3 w_j + 3\sum_{jk} v_j^2 v_k w_k \\
&= (C_4-3) \sum_j v_j^3 w_j + 3\norm{\bv}_2^2 (\bv^T\bw) \\
&= 3 (\bv^T\bw).
\end{split}
\end{equation}
Again, this holds for arbitrary unit vectors $\bv$ and $\bw$, which need not be $\mu$-flat.  When we choose $\bv$ and $\bw$ that are orthogonal, this bound serves the same purpose as equation (\ref{eqn-momx3v}).  

Following these changes, the rest of the proof goes through as before.


\section{Discussion}
\label{sec-discuss}

Broadly speaking, in this paper we have investigated the power of subgaussian measurements for phase retrieval.  It is known that for certain natural classes of subgaussian measurements, such as Bernoulli random vectors, phase retrieval is impossible, in that certain signals can never be distinguished without unambiguity.  We have shown that a large class of signals, namely those that are ``non-peaky'' (more precisely, those that are $\mu$-flat), can still be recovered in this setting.  This $\mu$-flatness condition, where $\mu$ can be a constant independent of the dimension $n$, is surprisingly weak.  We have extended recent results on stable uniqueness \cite{EM} and PhaseLift \cite{CL} to this setting.  

In addition, we have shown that for one particular example of a subgaussian measurement distribution, namely Bernoulli random vectors with erasures, one does not need any $\mu$-flatness restriction at all:  PhaseLift can recover \textit{all} vectors in $\RR^n$, using $m = O(n)$ measurements.  This is close to the information-theoretic lower-bound.

Our proof is based on a dual certificate argument, and it is an interesting question whether similar or better results could be derived using Mendelson's small ball method \cite{kueng2014}.  Indeed, the small ball method is known to yield stronger results in other scenarios, especially when the measurements are affected by noise. However, it is not a priori clear how to incorporate our non-peakiness condition into the framework of the small ball method. The small ball method analyzes the measurements corresponding to a matrix $\bX$ that lies in the descent cone of the nuclear norm, rather than the tangent space. Thus $\bX$ need not be of the form $\bx\bx_0^T + \bx_0\bx^T$, and it may have rank greater than two. Hence it is difficult to even precisely capture the $\mu$-flatness assumption in this framework. In our opinion, formulating and applying the small ball method for non-peaky signals is a very interesting direction for follow-up work. With such an approach, it may even be possible to generalize our strategy to measurements with non-vanishing mean, such as 0/1 Bernoulli measurements, as has been successfully done in the context of sparse recovery \cite{JK16}.

It would also be interesting to understand better what conditions on the measurements $\ba_i$ and the signal $\bx$ are necessary and sufficient for phase retrieval.  For example, Eldar and Mendelson's small-ball and fourth-moment conditions involve the $\ba_i$, while our $\mu$-flatness condition involves $\bx$.  Both of these seem to be special cases of some more general conditions that involve the $\ba_i$ and $\bx$ jointly.  In particular, these conditions seem to involve certain fourth moments of the $\ba_i$ projected onto directions that depend on $\bx$, as seen in equations (\ref{eqn-platypus}), (\ref{eqn-momx}), (\ref{eqn-momxv}) and (\ref{eqn-momx3v}).  Understanding these conditions may be helpful for generalizing our results to other types of measurements, and other classes of signals.


\vskip 11pt
\noindent
\textbf{Acknowledgements:}  Our work on this paper was stimulated by the Oberwolfach mini-workshop {\em Mathematical Physics meets Sparse Recovery} in April 2014, and in parts performed during the ICERM semester program {\em High-dimensional Approximation} in September 2014.  We thank Shahar Mendelson and the anonymous reviewers of the SampTA 2015 conference for several helpful comments, and Richard Kueng for his suggestion to consider random Bernoulli measurements with erasures.  Finally, we thank the reviewers of this journal for a number of insightful comments and perspectives.

Felix Krahmer's work on this topic was supported by the German Science Foundation (DFG) in the context of the Emmy Noether Junior Research Group KR4512/1-1 (RaSenQuaSI) and Project KR4512/2-1 as well as by the German Israeli Foundation (Grant no.~1266). Contributions to this work by NIST, an agency of the US government, are not subject to US copyright law.


\appendix

\section{Large-deviation bounds for random vectors}

The following variants of the Hoeffding and Bernstein inequalities, for sums of independent subgaussian and subexponential random vectors, are straight-forward generalizations of the standard results. 

\begin{lemma}[Vector Hoeffding inequality]
\label{lem-vector-hoeffding}
Let $\bx_1,\ldots,\bx_N$ be independent, centered, subgaussian random variables taking values in $\RR^n$, and let $K = \max_{i\in[N]} \norm{\bx_i}_{\psi_2}$.  Fix some $\ba = (a_1,\ldots,a_N) \in \RR^N$.  Let $0 < \eps < 1$ and $t \geq 0$.  Then 
\begin{equation}
\Pr\bigl[ \lVert \sum_{i=1}^N a_i\bx_i \rVert_2 \geq \tfrac{1}{1-\eps} t \bigr] 
\leq e\cdot \exp\bigl( n\ln(\tfrac{3}{\eps}) - \tfrac{ct^2}{K^2 \norm{\ba}_2^2} \bigr), 
\end{equation}
where $c>0$ is a universal constant.
\end{lemma}

\noindent
Proof:  We use the Hoeffding inequality for scalar random variables, together with a covering argument over the unit sphere.  

For any vector $\bw \in S^{n-1}$, we have $\norm{\bw^T\bx_i}_{\psi_2} \leq K$.  Using the scalar Hoeffding inequality (Prop. 5.10 in \cite{vershynin}), we get that 
\begin{equation}
\Pr\bigl[ \lvert \bw^T \sum_{i=1}^N a_i\bx_i \rvert \geq t \bigr] 
\leq e\cdot \exp\bigl( -\tfrac{ct^2}{K^2 \norm{\ba}_2^2} \bigr).
\end{equation}

Using Lemma 5.2 in \cite{vershynin}, we know that there is an $\eps$-net for $S^{n-1}$, with respect to the $\ell_2$ norm, with cardinality $(1+\tfrac{2}{\eps})^n$; call this set $\calN_\eps$.  Note that when $0 < \eps < 1$, we can simplify this to get $\abs{\calN_\eps} \leq (\tfrac{3}{\eps})^n$.  Taking the union bound over all $\bw \in \calN_\eps$, we get that 
\begin{multline}
\label{eqn-algae}
\Pr\bigl[ \exists \bw \in \calN_\eps \text{ s.t. } \lvert \bw^T \sum_{i=1}^N a_i\bx_i \rvert \geq t \bigr] \\
\leq e\cdot \exp\bigl( n\ln(\tfrac{3}{\eps}) - \tfrac{ct^2}{K^2 \norm{\ba}_2^2} \bigr).
\end{multline}

Finally, a standard argument (similar to Lemma 5.3 in \cite{vershynin}) shows that, for any vector $\bv \in \RR^n$, 
\begin{equation}
\max_{\bx \in \calN_\eps} \bv^T\bx \leq \norm{\bv}_2 
\leq \tfrac{1}{1-\eps} \max_{\bx \in \calN_\eps} \bv^T\bx.
\end{equation}
Thus, if $\lVert \sum_{i=1}^N a_i\bx_i \rVert_2 \geq \tfrac{1}{1-\eps} t$, then there must exist some $\bw \in \calN_\eps$ such that $\lvert \bw^T \sum_{i=1}^N a_i\bx_i \rvert \geq t$.  Combining this with equation (\ref{eqn-algae}) completes the proof.  $\square$

\vskip 11pt

\begin{lemma}[Vector Bernstein inequality]
\label{lem-vector-bernstein}
Let $\bx_1,\ldots,\bx_N$ be independent, centered, sub-exponential random variables taking values in $\RR^n$, and let $K = \max_{i\in[N]} \norm{\bx_i}_{\psi_1}$.  Fix some $\ba = (a_1,\ldots,a_N) \in \RR^N$.  Let $0 < \eps < 1$ and $t \geq 0$.  Then 
\begin{multline}
\Pr\bigl[ \lVert \sum_{i=1}^N a_i\bx_i \rVert_2 \geq \tfrac{1}{1-\eps} t \bigr] \\
\leq 2\exp\bigl( n\ln(\tfrac{3}{\eps}) - c\min\set{ \tfrac{t^2}{K^2 \norm{\ba}_2^2}, \, \tfrac{t}{K\norm{\ba}_\infty} } \bigr), 
\end{multline}
where $c>0$ is a universal constant.
\end{lemma}

\noindent
Proof:  We use the same argument as for Lemma \ref{lem-vector-hoeffding}, but starting from a scalar Bernstein inequality (Prop. 5.16 in \cite{vershynin}).  $\square$


\begin{IEEEbiographynophoto}{Felix Krahmer}
received his BSc in Mathematics from International University Bremen and his MSc as well as his PhD in Mathematics from New York
University under the supervision of Percy Deift and Sinan G\"unt\"urk. He was a postdoctoral fellow in the group of Holger Rauhut at the University of Bonn,
Germany, from 2009-2012. In 2012 he joined the University of G\"ottingen as an assistant professor for mathematical data analysis, where he has been awarded
an Emmy Noether Junior Research Group. Since 2015 he has been a tenure track assistant professor for optimization and data analysis in the department of
mathematics at the Technical University of Munich. His research interests include mathematical signal and data processing, applied harmonic analysis, and
random matrix theory.
\end{IEEEbiographynophoto}

\begin{IEEEbiographynophoto}{Yi-Kai Liu}
received his bachelors degree in Mathematics at Princeton University in 2002. He received his PhD in Computer Science at the University of California, San Diego, in 2007. He was the recipient of an NSF Mathematical Sciences Postdoctoral Research Fellowship, and a postdoctoral researcher at Caltech and at the University of California, Berkeley. In 2011 he joined the US National Institute of Standards and Technology (NIST). In addition, he is currently a Fellow at the Joint Center for Quantum Information and Computer Science (QuICS) at the University of Maryland. His research interests include quantum computation, cryptography, and machine learning.
\end{IEEEbiographynophoto}

\end{document}